\documentclass[prb, twocolumn,showpacs,preprintnumbers,amsmath,amssymb]{revtex4-1}
\usepackage{graphicx}
\usepackage[citecolor=blue,colorlinks=true,linkcolor=blue]{hyperref}
\usepackage{color}
\usepackage[usenames,dvipsnames]{xcolor}

\begin{document}

\title{SLKMC-II study of self-diffusion of small Ni clusters on Ni (111) surface}

\author{Syed Islamuddin Shah}
\email{islamuddin@knights.ucf.edu}
\author{Giridhar Nandipati}
\email{giridhar.nandipati@ucf.edu}
\author{Abdelkader Kara}
\email{abdelkader.kara@ucf.edu}
\author{Talat S. Rahman}
\email{talat.rahman@ucf.edu}
\affiliation{Department of Physics, University of Central Florida,  Orlando, FL  32816}

\date{\today}

\begin{abstract}

We studied self-diffusion of small 2D Ni islands (consisting of up to $10$ atoms) on Ni (111) surface using a self-learning kinetic Monte Carlo (SLKMC-II) method with an improved pattern-recognition scheme that allows inclusion  of both fcc and hcp sites in the simulations. In an SLKMC simulation, a database holds information about the local neighborhood of an atom and associated processes that is accumulated on-the-fly as the simulation proceeds. In this study, these diffusion processes were identified using the drag method, and their activation barriers calculated using a semi-empirical interaction potential based on the embedded-atom method. Although a variety of concerted, multi-atom and single-atom processes were automatically revealed in our simulations, we found that these small islands diffuse primarily via concerted diffusion processes.  We report diffusion coefficients for each island size at various tepmratures, the effective energy barrier for islands of each size and the processes most responsible for diffusion of islands of various sizes, including concerted and multi-atom processes that are not accessible under SLKMC-I or in short time-scale MD simulations.

\end{abstract}
\maketitle
\section{Introduction}
Surface diffusion is of interest not only because it is so different from diffusion in bulk solids \cite{diffbook} but because diffusion of adatoms on metal surfaces, individually or as a group via multi-atom or concerted diffusion processes plays an essential role in a wide variety of such surface phenomena as heterogenous catalysis, epitaxial crystal growth, surface reconstruction, phase transitions, segregation, and sintering.\cite{PAS} A precise knowledge of diffusion mechanisms is essential for understanding and control of these phenomena.\cite{Kax}  
Adatoms can diffuse on a substrate in a variety of ways, and competition between various types of diffusion processes (due to the differences in their rates) determines the shapes of the islands formed and (on macroscopic times scales), the morphological evolution of thin films. Hence a great deal of effort has been devoted to investigation of self-diffusion of adatom islands on metal surfaces, initially using field ion microscopy (FIM),\cite{bassett, tsong, wangss, wangprl1, wangprl2, kellog} and more recently scanning tunneling microscopy (STM).\cite{wen,pai1,giesenss,giesenprl,fern,vander,busse,muller,liu}  Because of inherent differences in the microscopic processes responsible for island diffusion on different metal surfaces, this is still an on-going research problem.  Both experimental and theoretical studies for various systems have succeeded in finding the activation barriers and prefactors for a single-adatom diffusion processes \cite{Ni1,Ni2,Ni3,Ni4,Ni5,Ni6,Ni7,Ni8,Ni9,Ni10,Ni11,Ni12,Ni13}. Ref~\onlinecite{diffbook} provides a good survey of those efforts. To the best of our knowledge, however, there has so far been no systematic experimental or theoretical effort to identify the diffusion mechanisms responsible for diffusion of small 2D Ni islands on Ni(111) and to calculate their activation barriers. In this article we report our results of doing so for such islands, ranging in size from 1 to 10 atoms.

Arrangement of atoms in the substrate of an fcc(111) surface results in two types of three-fold hollow sites for an adatom: the regular fcc site (with no atom beneath it in the second layer), and an hcp site (with an atom beneath it in the second layer). Occupancy of adatoms at fcc sites maintains the crystal stacking order (ABC stacking) of fcc structure, while occupancy of hcp sites leads to a stacking fault.   
Depending on its relative occupation energy, which is material dependent, an adatom can occupy one or the other of these sites. Which site is preferred on the  fcc(111) surface affects the way diffusion and hence growth proceeds.  It is therefore important to understand whether the diffusion proceeds via movement of atoms from fcc-to-fcc or hcp-to-hcp or fcc-to-hcp hcp-to-fcc sites. It has been observed experimetally that for smaller clusters mixed occupancy \cite{repp} of fcc \& hcp sites is possible.

A host of studies has been devoted to problems of self-diffusion and diffusion mechanisms on metal fcc(111) surfaces, almost exclusively, however, with either a preconceived set of processes or merely approximate activation barriers. It is nevertheless crucial to discover the full range of processes at work and to accurately establish the activation barrier of each. It is also well known that the fcc(111) surface, being atomically flat, has the least corrugated potential energy surface of any fcc surface, resulting in low diffusion barriers even for clusters to diffuse as a whole. Consequently, studies of diffusion processes on fcc(111) surfaces is a challenging problem for both experiment and simulation even to this day.
For a monomer and smaller islands like dimer, trimer and up to certain extent, tetramer, all possible diffusion processes may be guessed.  But as islands further increase in size, it becomes more difficult to enumerate all possible diffusion processes {\it a priori}. An  alternative is to resort to molecular dynamics (MD) simulation. But because diffusion processes are rare events, an MD simulation  cannot capture every microscopic process possible, as most of the computational time is spent in simulating atomic vibration of atoms.  Instead, to do a systematic study of small Ni island diffusion on Ni (111) surface we resorted to an on-lattice self-learning kinetic Monte Carlo (SLKMC-II) method, which enables us to study longer time-scales than are feasible with MD yet to find all the relevant atomic processes and their activation barriers on-the-fly, as KMC methods limited to {\it a priori} set of processes cannot do.
Moreover, whereas previous studies have used an on-lattice SLKMC method, \cite{slkmc1,slkmc2,pslkmc,iss_short} in which adatoms were restricted to fcc occupancy, in the present study both fcc and hcp occupancies are allowed, and are detectible by our recently  developed improved pattern-recognition scheme\cite{slkmcII}. 

The remainder of the paper is organized as follows. In Section~\ref{sim_details} we discuss the details of our SLKMC-II simulations, with particular attention to the way we find diffusion processes and calculate their activation barriers. In Section~\ref{results} we present details of concerted, important multi-atom and single-atom diffusion processes responsible for the diffusion of Ni islands as a function of island size. In Section~\ref{Dcoefficients} we present a quantitative analysis of diffusion coefficients at various temperatures and of effective energy barriers as a function of island size. In Section~\ref{discussion} we present our conclusions.

\section{Simulation Details}\label{sim_details}
To study Ni island diffusion on fcc Ni(111) surface, we carried out SLKMC simulations using the pattern-recognition scheme we developed recently \cite{slkmcII} that includes both fcc and hcp sites in the identification of an atom's neighborhood. 
Various types of diffusion processes are possible, and their activation barrier depends on the atom's local neighborhood. Whenever a new neighborhood around an atom is identified, a saddle-point search is carried out to find all the possible atomic processes and calculate their activation barriers -- provided that it has at least one similar empty site in the second ring\cite{slkmcII}, since when an atom occupies an fcc (or alternatively an hcp) site, the nearest neighbor (NN) hcp (or, correspondingly, fcc) sites cannot be occupied. In our simulations we used a system size of 16x16x5 with the bottom 2 layers fixed, and carried out saddle-point searches using the drag method. In this method a central or active atom is dragged  in small steps towards a probable final position. If the central atom is on an fcc (hcp) site, then it is dragged towards a NN vacant fcc (hcp) site in the second ring. Since atoms are allowed to occupy either hcp or fcc sites, an atom being dragged from an fcc (hcp) site to a neighboring similar site is allowed to relax to an intermediate hcp (fcc) site in between the two fcc (hcp) sites. In other words, processes are possible in which atoms in an island may occupy fcc, hcp or both fcc \& hcp sites simultaneously. In the drag method, the atom being dragged is always constrained in the direction of the reaction coordinate but allowed to relax along its other degrees of freedom (those perpendicular to the reaction coordinate), while all the other atoms in the system are allowed to relax in all degrees of freedom. Once the transition state is found, the entire system is completely relaxed to find the final state of the process. The activation barrier of the process is the difference between the energies of the transition and initial states. We have verified the activation barriers of some of the key processes found by the drag method using the (more accurate but computationaly expensive) nudged-elastic band (NEB) method\cite{neb}, and found no significant difference. For inter-atomic interactions, we used an interaction potential based on the embedded-atom method (EAM) as developed by Foiles {\it et al}.\cite{Foiles}. In all our SLKMC simulations we used the same pre-exponential factor of $10^{12}$s$^{-1}$, which has been demonstrated to be a good assumption for such systems as the one under examination here.\cite{handan-prb, handan-ss}
 
For the small islands under study here (1-10 atoms), we found that when an atom is dragged rest of the atoms in the island usually follow. For very small islands (1-4 atoms), $\it all$  of the processes identified by the drag method were concerted-diffusion processes. As island size increases we found single-atom and multi-atom processes as well. For islands of size 5-6, even single-atom detachment processes are identified and stored in the database (even though they are not allowed in our simulations). 
To account for all types of processes associated with both compact and non-compact shapes -- especially concerted processes and multi-atom processes -- we used $10$ rings to identify the neighborhood around an active atom in our SLKMC simulations. Using $10$ rings corresponds to including fifth nearest-neighbor interactions. To make sure we identified all the single-atom processes, we also carried out saddle-point searches with all of the atoms fixed except the atom being dragged. Although there is no infallible method for discovering all possible processes, we did exhaust the search for possible processes identifiable using the drag method.

In order to save computational time, we first carried out SLKMC simulations at 700K for each island size, and used the database thus generated to carry out our simulations for the same size at lower tempratures (300, 400, 500 and 600K ). The rationale for this approach is that an island goes through many more shapes at higher tempratures: when a simulation is carried out at a lower temprature starting out with a database generated at a higher temprature, it only rarely finds an unknown configuration. It is not possible, however, to economize on computational time by using, for the smaller islands under study here, a database generated for (say) the larger among them, because the types of processes possible (along with their respective barriers) are dependent on an island's particular size.

\section{Results}\label{results}

As mentioned above,  all of the processes for a given island are identified and their activation barriers calculated, and stored in a database on-the-fly. We discuss in this section, however, only key processes of the various general types (concerted, multi-atom and single-atom).

\begin{figure}
\center{\includegraphics [width=5.5cm]{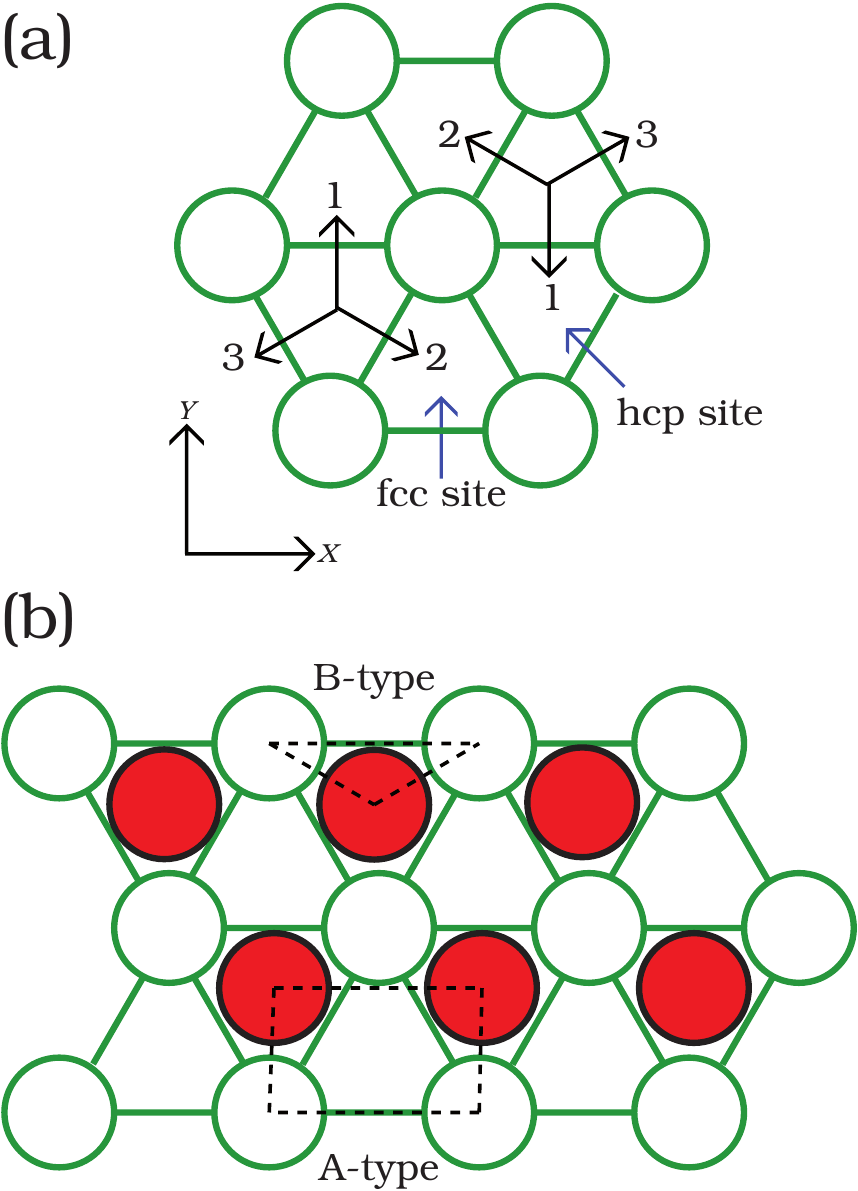} 
\caption{\label{monomer}{(a) fcc and hcp sites on an fcc(111) surface, with corresponding directions for concerted diffusion processes; (b) A-type and B-type step edges (here, for an all-hcp island) for the same surface.}}}
\end{figure}

Fig.~\ref{monomer}(a) is a sketch of the fcc(111) surface with its adsorption sites marked as fcc and hcp. Determining whether an adatom is on an fcc or on an hcp site on this surface requires knowledge of at least $2$ substrate layers below the adatom layer. In all our figures we show only the adatom layer and the layer below (the top substrate layer) with the convention that the center of an upward-pointing triangle (along the y-axis) formed by the (top layer) substrate atoms is an fcc site, while the center of  a  downward-pointing triangle pointing triangle is an hcp site.  An island on an fcc (111) surface can be on fcc sites or on hcp sites or a combination of both sites (some atoms of the island sitting on fcc sites and the rest on hcp sites).  Depending on the type of material either the fcc or the hcp site will be energetically favorable. As we shall see for each island size under study here, the fcc site for Ni(111) is always at least slightly more favorable than the hcp site.  

A compact adatom island on an fcc(111) surface can move in the three directions shown in Fig.\ref{monomer}(a). Note that the numbering scheme for the directions open to an atom on an fcc site is inverse to that for those open to an atom on an hcp site (see Fig.~\ref{monomer}(a)). We follow the enumeration convention for directions distinguished in Fig.~\ref{monomer}(a) throughout the article in tabulating activation barriers for concerted processes for islands of various sizes and shapes. Concerted processes involve all atoms moving together from all-fcc sites to all-hcp sites or vice-versa. In a concerted diffusion process a cluster can either translate in one of the three directions shown in Fig.~\ref{monomer} (concerted translation) or rotate around an axis (around the center of mass), either clockwise or anti-clockwise (concerted rotation). Since concerted rotational processes do not produce any displacement in the center of mass of an island, they do not contribute to island diffusion. Depending on the size of the island  and its shape, activation barriers for the processes  in these three directions can be different.

Activation barriers for single-atom processes, however, depend on the type of step-edge along which atom diffuses. Fig.~\ref{monomer}(b) shows, using the example of a 6-atom hcp island, how an A-type step-edge $\--$ a (100) micro-step differs from a B-type step-edge $\--$ a (111) micro-step. We discuss important single-atom diffusion processes systematically and in detail in Sub-section~\ref{single}.

As island size increases not only does the frequency of single-atom processes increase but the frequency of multi-atom processes does so as well. All multi-atom mechanisms involve shearing. A special case is reptation mechanism\cite{Chirita1,Chirita2}, a two-step shearing process that moves the cluster from all-fcc to all-hcp sites or the reverse: first, part of the island moves from fcc to hcp sites; then the rest of the island moves from fcc to hcp. Hence at the intermediate stage, the island has mixed fcc-hcp occupancy. In case of Ni-island diffusion, reptation processes occur only when the shape of the island becomes non-compact.  We will discuss reptation in detail when we take up islands of size 8-10.

\subsection{Monomer}

As mentioned earlier, much work has been done to determine activation barrier for Ni monomer diffusion on Ni (111) surface\cite{diffbook}. A monomer on fcc (111) surface can adsorb either on an fcc or an hcp site. We find that adsorption of an adatom on an fcc site is slightly favored over than on an hcp site by $0.002$ eV -- in good agreement with the value reported in Ref~\onlinecite{Ni4}.  Diffusion of a monomer occurs through hopping between fcc  sites via an intermediate hcp site. We find the activation barrier for a monomer's hopping from an fcc site to a neighboring hcp site to be $0.059$ eV while that for the reverse process is $0.057$ eV. The effective energy barrier for monomer is found to be 0.057 eV, which is consistent with the result reported by Liu {\it et al}.\cite{liu} of $0.056$ eV. 

\subsection{Dimer}

\begin{figure}
\center{\includegraphics [width=7.0cm]{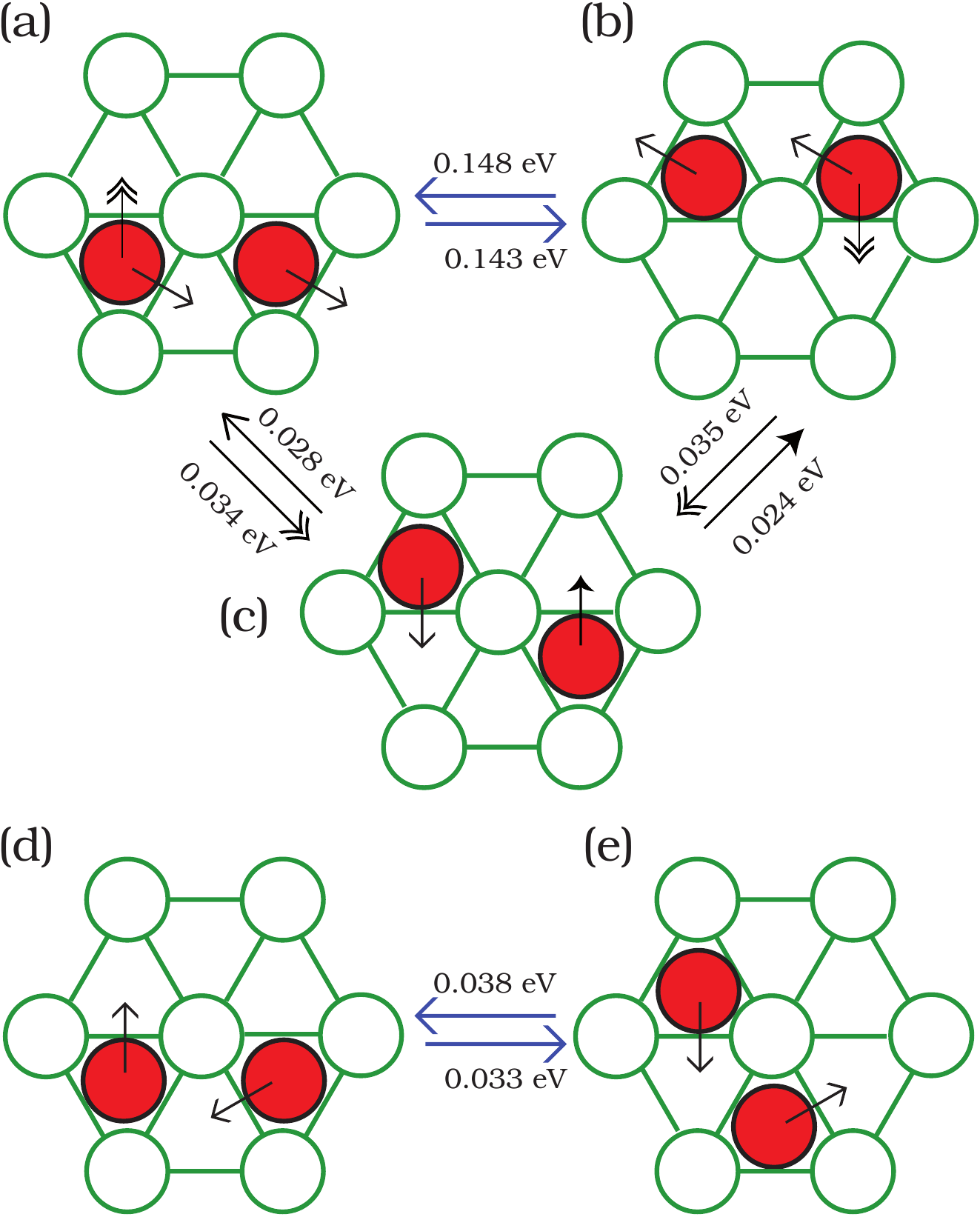}
\caption{\label{dimer}{Possible configurations for a dimer, with activation barriers for concerted diffusion processes. (a) FF dimer (both atoms on fcc sites); (b) HH dimer (both atoms on hcp sites); (c) FH dimer (one atom on an fcc and the other on an hcp site); (d) FF dimer in concerted clockwise rotation and (e) HH dimer in concerted counter-clockwise rotation.}}}
\end{figure}

On any fcc(111) surface a dimer (of the same species) can have three possible arrangements: both atoms on fcc sites (an FF-dimer, Fig.~\ref{dimer}(b)), both on hcp sites (an HH-dimer, Fig.~\ref{dimer}(a)) or one atom on an fcc and the other on an hcp site (an FH-dimer,  Fig.~\ref{dimer}(c)). We find that the FF-dimer is energetically more favorable than the HH-dimer by $0.005$eV and the FH-dimer the least favorable by $0.011$ eV. We find that both FF and HH dimers diffuse via concerted as well as single-atom processes, whereas the FH-dimer diffuses via single-atom processes only. 

\begin{table}
\caption{\label{tdimer}Activation barriers (in eV) of concerted processes for dimer diffusion.}
\begin{tabular}{ c  c c c c c c }
\hline
\hline
Direction~~&fcc & hcp\\
\hline
\hline
1~~ & 0.071 ~& 0.066 ~\\
2~~ & 0.148 ~& 0.143 ~\\
3~~ & 0.148 ~& 0.143 ~\\
\hline
\hline
\end{tabular}
\end{table}

In concerted diffusion processes, both atoms in a FF (HH) dimer move from fcc (hcp) to the nearest hcp (fcc) sites as shown in Fig.~\ref{dimer} (a)\&(b) and Fig.~\ref{dimer} (d)\&(e), thereby converting an FF (HH) dimer into an HH (FF) dimer. In the case of an FF (HH) dimer, the activation barrier for concerted translational (Fig.~\ref{dimer} (a)\&(b)) is  $0.148$ eV ($0.143$ eV) while that for concerted rotation (Fig.~\ref{dimer} (d)\&(e)) is $0.038$ eV.  In concerted dimer rotation, the activation barriers for both clockwise and anti-clockwise directions are the same, as they are symmetric to each other. Activation barriers for translational concerted diffusion processes in all three directions (see Fig.~\ref{monomer}) both for FF and HH dimers are reported in Table.~\ref{tdimer}. Our results for concerted processes are $0.028$ eV higher than the corresponding activation barriers for a dimer reported in Ref. \onlinecite{liu}. (This difference $\--$ as with those in what follows $\--$ may be due to the different inter-atomic potential employed in their study and ours.)

Single-atom processes transform both FF and HH dimers into an FH-dimer. In this case one of the fcc atoms in an FF-dimer or an hcp atom in an HH-dimer moves to a nearest-neighbor hcp or an fcc site respectively, as shown in Fig.~\ref{dimer} (a)\&(b) with the double-headed arrow. The activation barriers are $0.034$ eV and $0.035$ eV for hcp and fcc dimer, respectively.
In the case of an FH-dimer, two types of single-atom diffusion processes are possible, as shown in Fig.~\ref{dimer}: an fcc atom moves to the nearest hcp site in the direction of the open arrowhead, forming an HH-dimer, or an hcp atom moves in the direction of the solid arrowhead to the nearest fcc site, forming an FF-dimer. The activation barriers for these processes are $0.028$ eV and $0.024$ eV, respectively. 

\subsection{Trimer}

\begin{figure}[ht]
\center{\includegraphics [width=8.5cm]{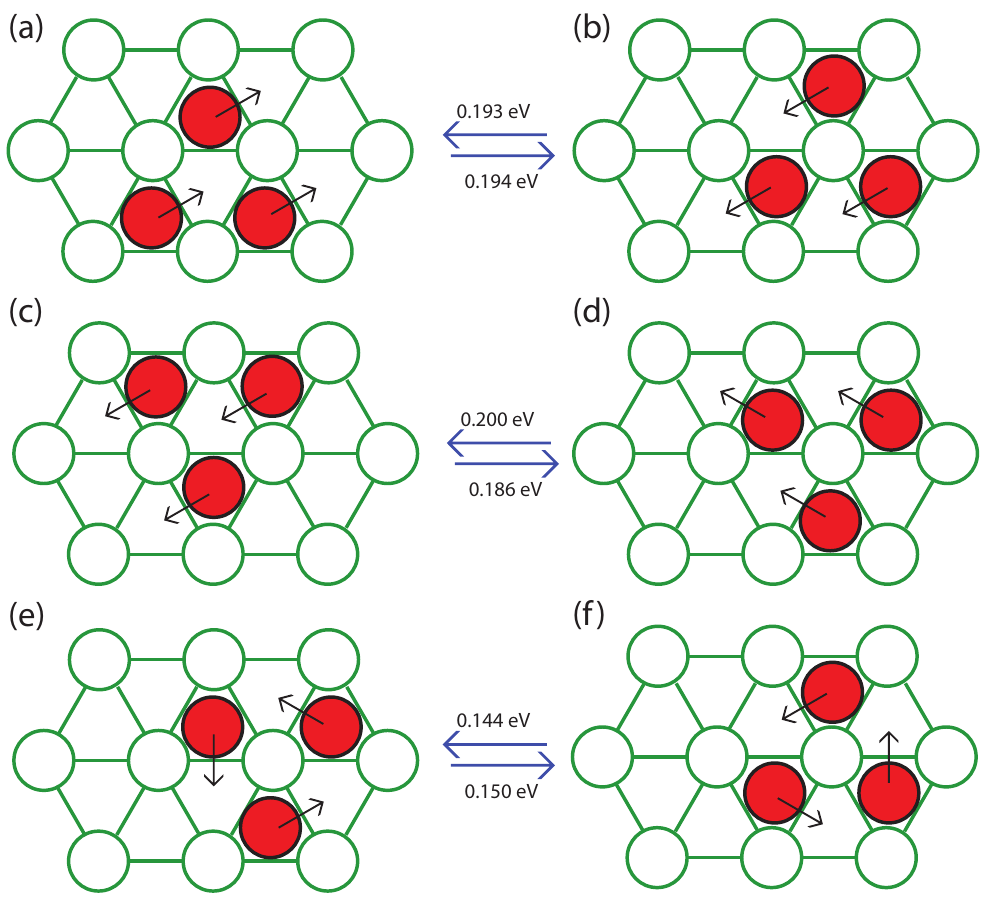} 
\caption{\label{trimer}{Possible arrangements of atoms in a trimer, with possible concerted diffusion processes and their activation barriers. (a)$\--$(d) Concerted translation: (a) F3H-all atoms on fcc sites centered around an hcp site; (b) H3T-all atoms on hcp sites centered around a top site; (c) H3F-all atoms on hcp sites centered around an fcc site; (d) F3T-all atoms on fcc sites centered around a top site. (e) \& (f) Concerted rotation: F3T and H3T respectively.}}}
\end{figure}

Depending on where a third atom is attached to the dimers shown in Fig.~\ref{dimer}(a \& b), there are four possible arrangements of atoms in a compact trimer: two types of fcc timers -- one centered around an hcp site (F3H), the other centered around a top site (F3T) (see Figs.~\ref{trimer}(a) \& (d)) $\--$ and two types of hcp trimers $\--$ one centered around an fcc site (H3F), the other centered around a top site (H3T) (Figs.~\ref{trimer}(c) \& (b)). Although all four trimers have the same shape, their local environment is different, so that their adsorption energies are distinct, as are the activation barriers for their possible diffusion processes. F3T timer is the most energetically favorable: F3H, H3T and H3F are less energetically favorable by $0.006$, $0.007$ and $0.0013$ eV, respectively. It should also be noted that although trimers can take on non-compact shapes, the configurations depicted in Fig.~\ref{trimer} are the most frequently observed in our trimer simulations.

In the case of F3T and H3T trimers, two types of concerted processes were observed, a non-diffusive concerted rotation (clockwise and anti-clockwise) (Fig~\ref{trimer} (e) \& (f)) and a diffusive concerted translation (in all three directions) (Fig~\ref{trimer} (d) \& (b)). Concerted rotation processes transform an H3T timer into a F3T trimer and vice versa.  The activation barrier for the concerted rotation processes for F3T trimer is $0.150$ eV while that for those of the H3T trimer is $0.144$ eV.  Translation transforms an F3T timer into a H3F timer and vice versa. The activation barrier for the concerted translations possible for these two trimers are $0.200$ eV and $0.193$ eV for F3T and H3T, respectively.  The activation barrier of $0.200$ eV for translational motion of F3T trimer is in agreement with the value reported in Ref~\onlinecite{liu}. For F3H and H3F trimers,  only concerted translation processes are possible; their activation barriers are $0.194$ eV and $0.186$ eV (the value reported for the same process in Ref~\onlinecite{liu} is $0.187$ eV), respectively. Fig.~\ref{trimer} (a-b) \& (c-d) reveal that these concerted diffusion processes transform an F3H into an H3T timer and an H3F to an F3T trimer.  Since the shape of these trimers is symmetric (see Fig ~\ref{trimer} (a) \& (c)), the activation barriers for their diffusion in all $3$ possible directions are the same.

\begin{table}
\caption{\label{ttrimer}Activation barriers (in eV) for single-atom diffusion processes for an H3T compact trimer in the directions shown in Fig.~\ref{trimer_single}. }
\begin{tabular}{ c  c c c c c c }
\hline	
\hline
Type~~&1&2&3&4\\
\hline
\hline
F3T~~ & 0.439 ~& 0.858 ~& 0.858 ~& 0.439 \\
F3H~~ & 0.432 ~& 0.875 ~& 0.875 ~& 0.432 \\
H3T~~ & 0.436 ~& 0.856 ~& 0.856 ~& 0.436 \\
H3F ~~& 0.429 ~& 0.872 ~& 0.872 ~& 0.429 \\
\hline
\hline
\end{tabular}
\end{table}

\begin{figure}[ht]
\center{\includegraphics [width=4.0cm]{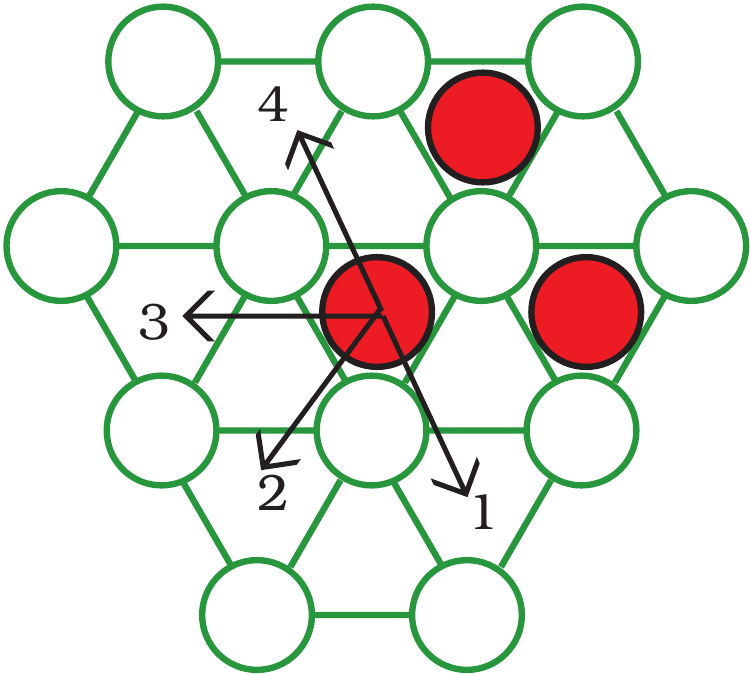} 
\caption{\label{trimer_single}{Single-atom processes possible for an H3T trimer. Activation barriers for the processes in these 4 directions for the 4 posible trimer configurations are given in Table.~\ref{ttrimer}}}}
\end{figure}

As for single-atom processes in the case of a trimer:  an atom can move in $4$ different directions as shown in Fig.~\ref{trimer_single}, resulting $2$ different types of single-atom processes: directions $1$ \& $4$ correspond to edge-diffusion processes which open up the trimer; directions $2$ \& $3$ correspond to detachment processes (excluded from the present study, which is confined to diffusion of single whole islands, in which an island's integrity [and hence its size] is maintained).  We note that these processes move atoms from fcc (hcp) to nearest fcc (hcp) site.  Activation barriers for processes in these $4$ directions for different types of trimers are given in Table.~\ref{ttrimer}. Because these activation barriers are so high relative to those for concerted processes, single-atom processes were rarely observed in our simulations of trimer diffusion. 

We note that as island size increases, possible types of single-atom processes increases as well (though with the decamer, basically all possible types have appeared). Accordingly, it is convenient to defer detailed discussion of single-atom processes until later (Sub-section.~\ref{single})

\subsection{Tetramer}

\begin{figure}[ht]
\center{\includegraphics [width=8.5cm]{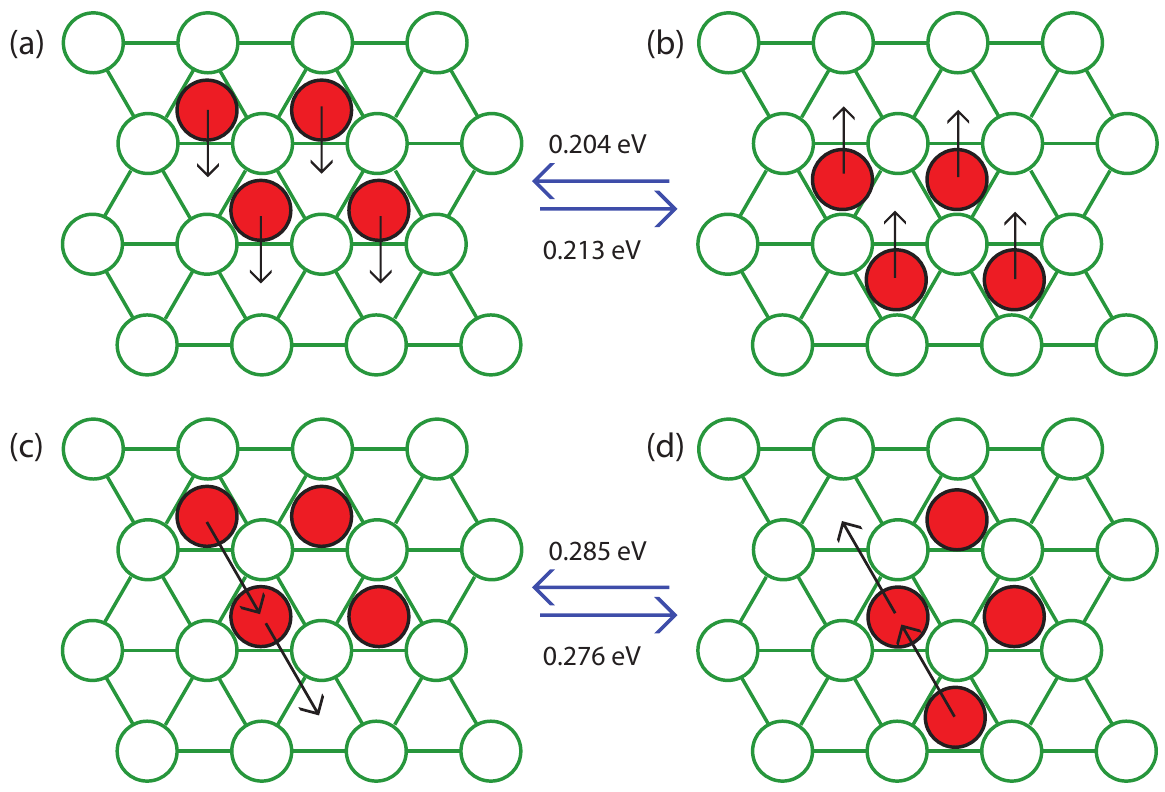} 
\caption{\label{tetramer}{Diffusion processes possible for a tetramer: (a) \& (b) concerted diffusion along the short diagonal; (c) \& (d) shearing processes.}}}
\end{figure}

Adding another atom to any of the trimers shown in Fig.~\ref{trimer} (a - d) results in the formation of a compact tetramer, diamond-shaped $\--$ with a long diagonal (along the line joining farthest atoms) and a short one  (perpendicular to the long one), as shown in Fig.~\ref{tetramer}. Once again the fcc island (Fig.~\ref{tetramer}(a)) is energetically more favorable than the hcp one $\--$ in this case, by $0.009$ eV. Three types of translational concerted diffusion processes are possible for each of the fcc and hcp tetramers, that is, one along each of the three directions specified in Fig.\ref{monomer}. An example of a cncerted fcc-to-hcp process (along direction 1) for a tetramer is shown in Fig.~\ref{tetramer}(a); its activation barrier is 0.213 eV. The reverse process (hcp to fcc) is shown in Fig.~\ref{tetramer}(b); its activation barrier is 0.204 eV (the value reported in Ref. 18 is 0.210 eV). Because the process in direction 3 is symmetric to that in direction 1, the energy barriers for these processes are identical, as are those for the reverse processes. The energy barrier along direction 2 is 0.313 eV from fcc to hcp and 0.304 eV from hcp to fcc. These values are systematically displayed in Table.~\ref{ttetramer}.

Although the multi-atom processes shown in Fig.~\ref{tetramer}(c \& d) have activation barriers lower than those of single-atom processes, they could not be found using the drag method. These processes thus had to be added manually to the database and their activation barriers obtained using the nudged elastic band method (NEB). In these multi-atom processes, two atoms move togehter in the same direction, the result is a shearing mechanism as shown in Fig.~\ref{tetramer}(c) \& (d). For this shearing process, from fcc to hcp, the activation barrier is 0.285 eV; that for the reverse process from hcp to fcc is 0.276 eV. The drag method also finds single-atom processes, but because in tetramers (as in islands of size 3 $\--$ 7) these have higher activation barriers than those of concerted processes, they were not observed during the simulations.

\begin{table}
\caption{\label{ttetramer}Activation barriers (eV) for the concerted tetramer translation processes shown in Fig.~\ref{tetramer}(a)\&(b). }
\begin{tabular}{ c  c c c c c c }
\hline
\hline
Direction~~&fcc &hcp\\
\hline
\hline
1~~ & 0.213 ~& 0.204 ~\\
2~~ & 0.313 ~& 0.304 ~\\
3~~ & 0.213 ~& 0.204 ~\\
\hline
\hline
\end{tabular}
\end{table}

\subsection{Pentamer}
\begin{figure}[ht]
\center{\includegraphics [width=7.0cm]{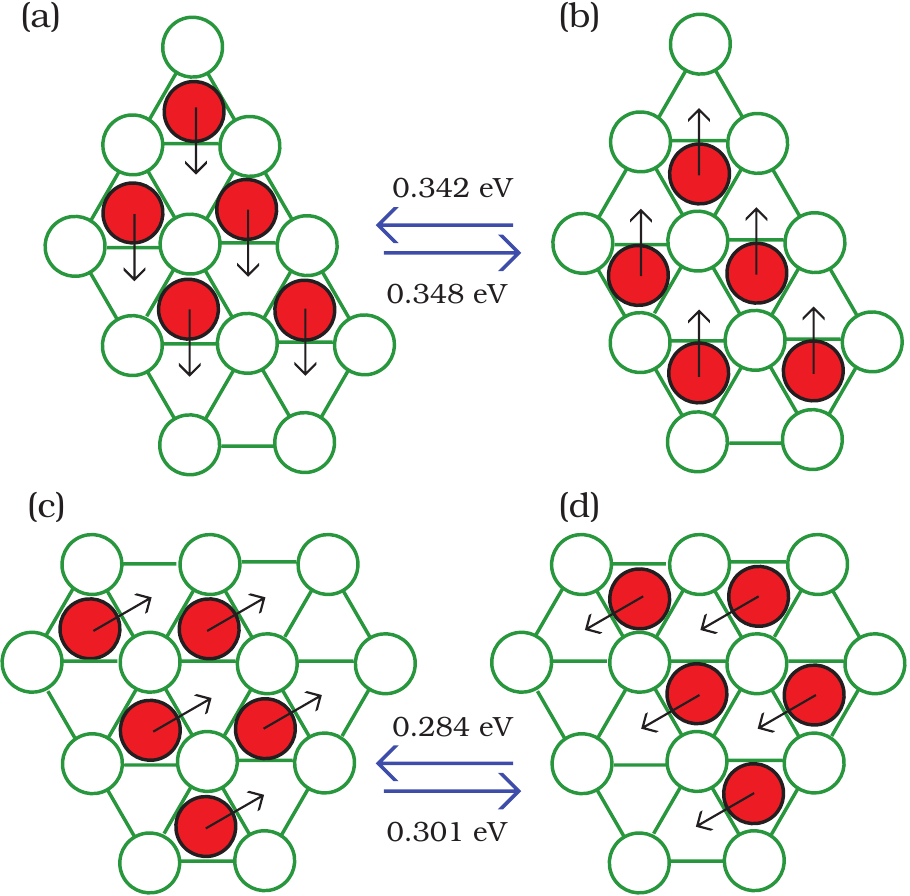} 
\caption{\label{pentamer-m}{Examples of concerted diffusion processes for a compact pentamer, along with their activation barriers.}}}
\end{figure}
The compact shapes of a pentamer can be obtained by attaching an atom to a diamond-shaped tetramer. Although the geometries of compact pentamer clusters thus obtained are the same, the island's diffusion is crucially affected by where this additional atom is placed: attachment of an atom to an A-type step-edge of an fcc tetramer results in the long A-type step-edge pentamer shown in Fig.~\ref{pentamer-m}(c); attachment of an atom to a B-type step-edge of the same tetramer results in the long B-type step-edge pentamer shown in Fig.~\ref{pentamer-m}(a)); the corresponding results of attaching an atom to an hcp tetramer are shown in Figs.~\ref{pentamer-m}(b) and (d), respectively. The most energetically favorable of these is the fcc pentamer with a long A-type step-edge (Fig.~\ref{pentamer-m}(c)); less favorlable by $0.005$  eV are the fcc pentamer with a long B-type step-edge (Fig.~\ref{pentamer-m}(a)), by $0.011$ eV the hcp pentamer with a long A-type step-edge, and by $0.017$ eV the hcp pentamer with a long B-type step-edge. That is: as usual, fcc islands are more stable than hcp ones. And, within each of those types, pentamers with a long A-type step-edge are more stable than those with a long B-type step-edge.
\begin{figure}[ht]
\center{\includegraphics [width=8.5cm]{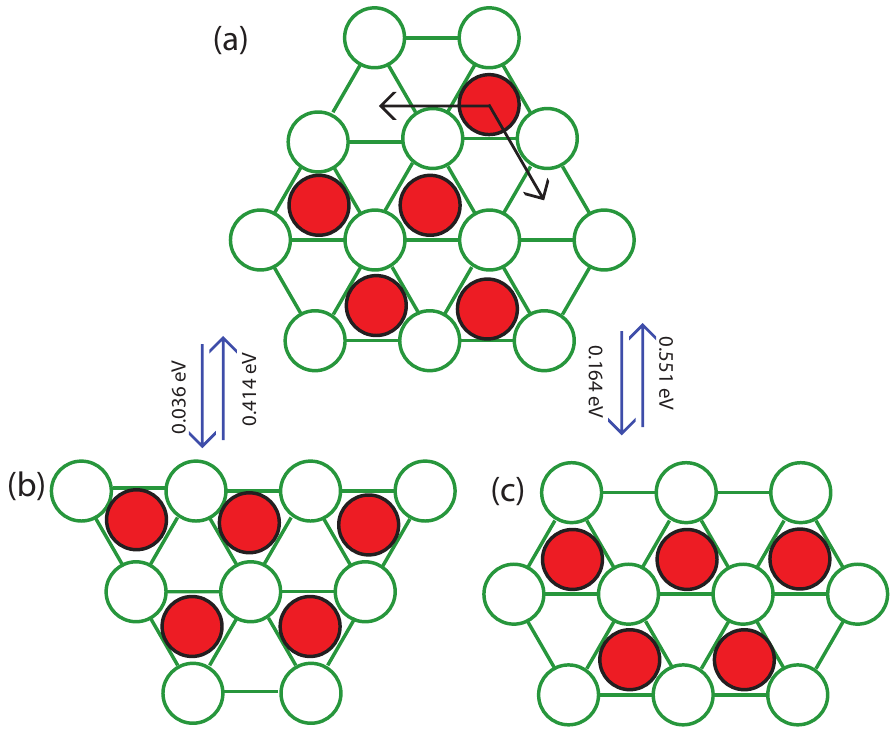} 
\caption{\label{pentamer-s}{The single-atom processes that convert the long A-type step-edge pentamer to the long B-type step-edge pentamer.}}}
\end{figure}
\begin{table}[ht]
\caption{\label{tpentamer}Activation barriers (eV) without parantheses are for concerted-translation processes of pentamers with a long A-type step-edge, as shown in Figs.~\ref{pentamer-m} (c)\&(b); barriers in parentheses are for such processes for pentamers with a long B-type step-edge, as shown in Figs.~\ref{pentamer-m} (a)\&(d).}
\begin{tabular}{ c  c c c c c c }
\hline
\hline
Directions~~&fcc & hcp\\
	    ~~&A       (B) &A      (B)\\
\hline
\hline
1~~ & 0.348 (0.301)~& 0.342 (0.284) ~\\
2~~ & 0.348 (0.353)~& 0.342 (0.337) ~\\
3~~ & 0.295 (0.353)~& 0.289 (0.337) ~\\
\hline
\hline
\end{tabular}
\end{table}

In our simulations we found that compact pentamers diffuse mostly via concerted diffusion processes, which displace the island as a whole from fcc-to-hcp or vice-versa. Fig.~\ref{pentamer-m} shows concerted diffusion processes along direction 1 for the long B-type step-edge pentamer and along direction 3 for the long A-type step-edge pentamer. Table~\ref{tpentamer} displays activation barriers for concerted processes in all 3 directions for both types of pentamer. Fig.~\ref{pentamer-s} (a),(b) \& (c) shows the single-atom processes that transform an fcc pentamer from long A-type (= short B-type) to a short A-type (= long B-type) cluster, with the activation barrier for each.
\subsection{Hexamer}
\begin{figure}[ht]
\center{\includegraphics [width=8.5cm]{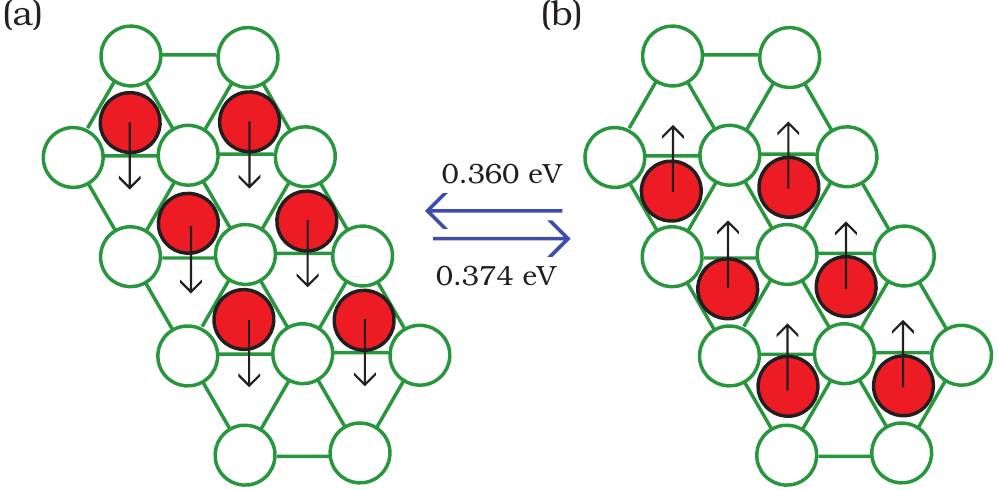} 
\caption{\label{hexamer-1}{Parallelogramic hexamers obtained by extending (by one atom) the shorter edge of either the long A-type or long B-type pentamers: (a) fcc cluster; (b) hcp cluster. The activation barriers indicated are for concerted diffusion in direction $1$.}}}
\end{figure}

\begin{figure}[ht]
\center{\includegraphics [width=8.5cm]{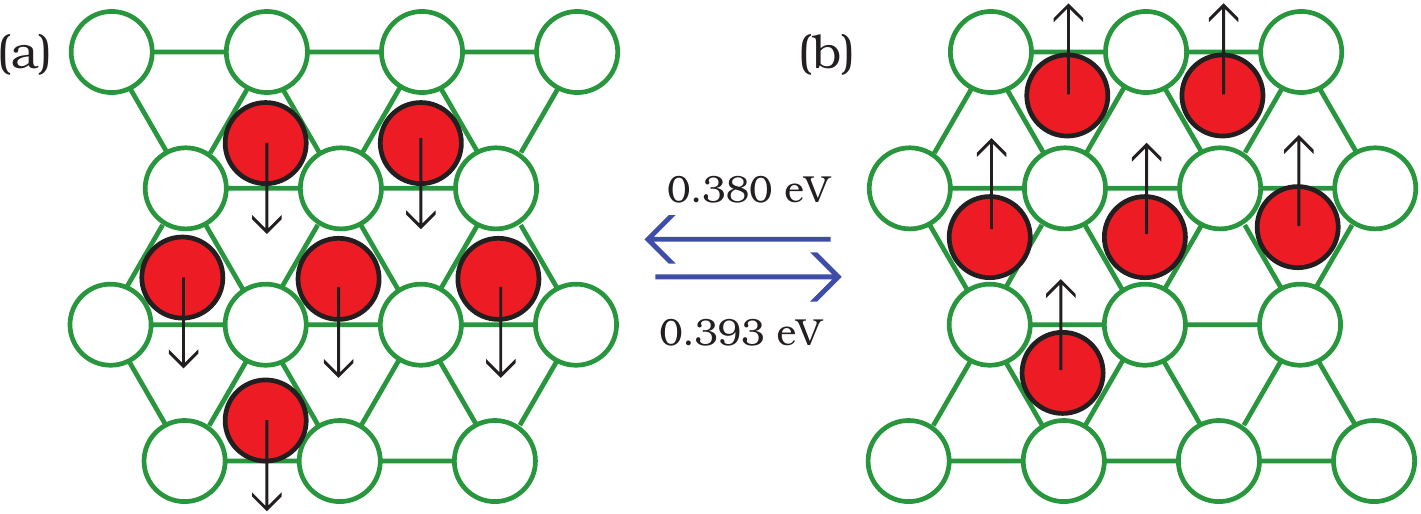} 
\caption{\label{hexamer-2}{Irregular hexamers obtained by attaching an atom to the long edge of a pentamer: (a) fcc cluster; (b) hcp cluster. The activation barriers indicated are for concerted diffusion in direction $1$.}}}
\end{figure}

\begin{figure}[ht]
\center{\includegraphics [width=8.5cm]{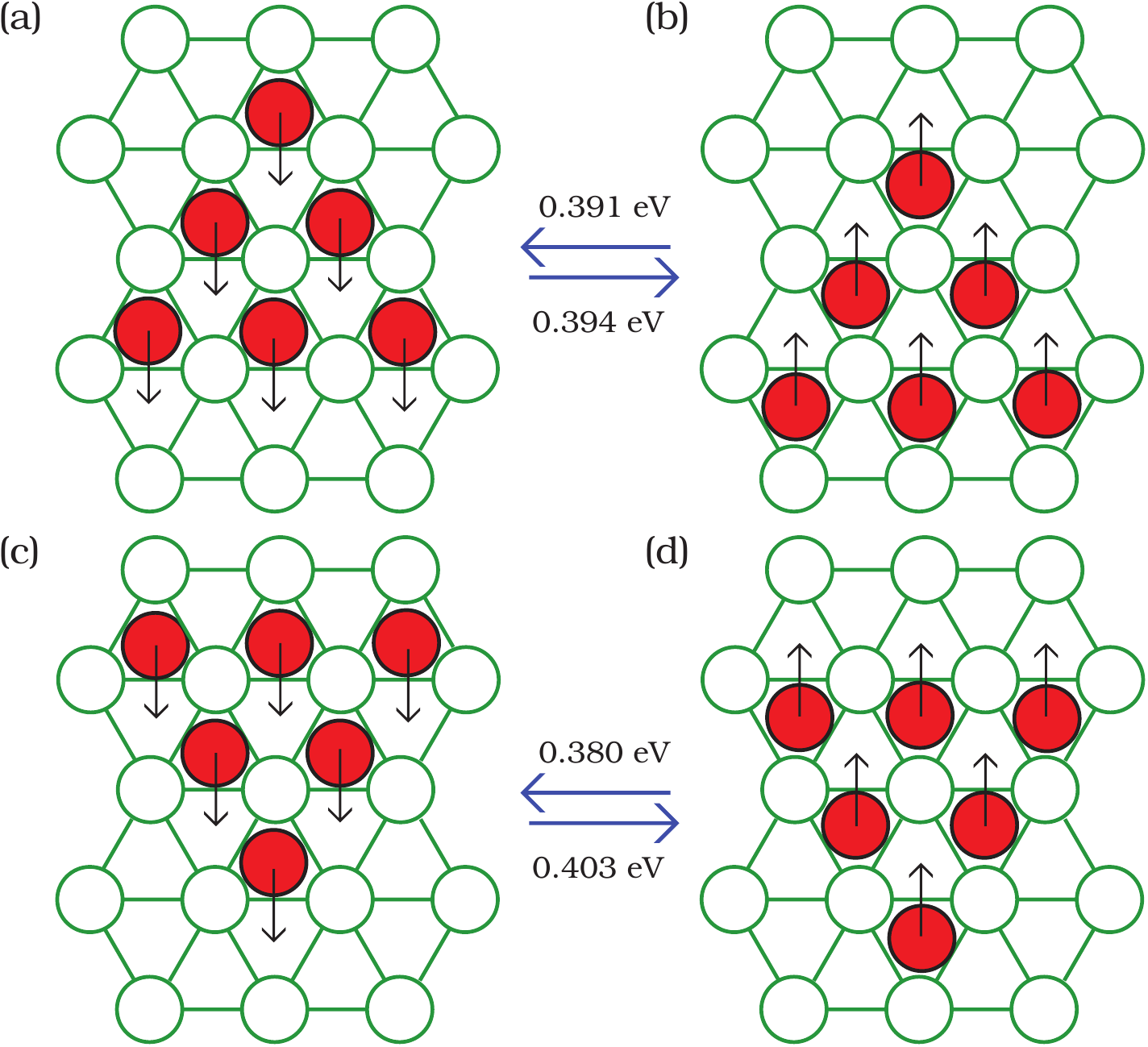} 
\caption{\label{hexamer-3}{Triangular hexamers obtained by adding an atom to the short edge of a pentamer: (a) fcc hexamer with B-type step edges; (b) hcp hexamer with A-type step edges; (c) fcc hexamer with A-type step edges; (d) hcp hexamer with B-type step edges. The activation barriers here are for concerted diffusion in direction $1$.}}}
\end{figure}

Depending on whether a sixth atom is attached to a long A-type or to a long B-type step-edge pentamer, there are $3$ possible compact shapes for a hexamer: (1) when an atom is added in such a way as to extend the  shorter edge of either a long A-type or a long B-type step-edge pentamer, the result is one of the parallelogramic hexamers shown in Fig.~\ref{hexamer-1}; (2) when an atom is attached to the long edge of either type of pentamer, the result is the one of the irregular hexamers shown in Fig.~\ref{hexamer-2}; ($3$) when an atom is added to the shorter edge of either type of pentamer, the result is one of the triangular hexamers with all step edges of either the A-type (Figs.~\ref{hexamer-3}(b) \&(c)) or of the B-type (Figs.~\ref{hexamer-3}(a) \&(d)).
\begin{table}[ht]
\caption{\label{hexamer-stabilities}Relative stabilities of the hexamers most frequently observed in our simulations. P = Parallelogram; I = Irregular; T = Triangular.}
\begin{tabular}{ l  c c c c }
\hline
\hline
Type~~&Shape ~~&Description ~~&Energy (eV)~~&Reference\\
\hline
\hline
%fcc~~ &P~&equal lenght \& no. of A \& B steps ~&0.0~&Fig.~\ref{hexamer-1}(a)~\\
fcc~~&P~&equal A \& B steps ~&0~&Fig.~\ref{hexamer-1}(a)~\\
fcc~~&I~&edge atom on A step ~&0.011~&Fig.~\ref{hexamer-2}(b)~\\
fcc~~&I~&edge atom on B step ~&0.011~&not shown~\\
%hcp~~&P~&equal length \& number of A \& B steps ~&0.014~&Fig.~\ref{hexamer-1}(b)~\\
hcp~~&P~&equal A \& B steps ~&0.014~&Fig.~\ref{hexamer-1}(b)~\\
fcc~~ &T~&all A steps ~&0.019~&Fig.~\ref{hexamer-3}(c)~\\
hcp~~ &I~&edge atom on B step ~&0.024~&Fig.~\ref{hexamer-2}(b)~\\
hcp~~ &I~&edge atom on A step ~&0.024~&not shown~\\
fcc~~ &T~&all B steps ~&0.029~&Fig.~\ref{hexamer-3}(a)~\\
hcp~~ &T~&all A steps ~&0.032~&Fig.~\ref{hexamer-3}(b)~\\
hcp~~ &T~&all B steps ~&0.042~&Fig.~\ref{hexamer-3}(d)~\\
\hline
\hline
\end{tabular}
\end{table}
Table.~\ref{hexamer-stabilities}  shows the order of relative stabilties of the hexamers most frequently obseved in our simulations. It reveals that, for $\it a$ $\it given$ $\it shape$, hexamers on fcc sites are more energetically favored than those on hcp sites, and that among hexamers on fcc sites (as for those on hcp sites), clusters in which A steps are longer than B steps are more stable.

Figs.~\ref{hexamer-1}-\ref{hexamer-3} also show concerted diffusion processes (in direction 1) for these hexamers, together with the activation barriers for each. Tables.~\ref{thexamer-1} \& \ref{thexamer-2} give activation barriers for the hexamers shown in Figs.~\ref{hexamer-1} \& \ref{hexamer-2}, respectively. Since triangular hexamers (Fig.~\ref{hexamer-3}) are symmetric, their activation barriers for concerted diffusion are same in all three directions.

\begin{table}[ht]
\caption{\label{thexamer-1}Activation barriers (eV) of the concerted translation processes in all three directions for the hexamer shown in Fig.\ref{hexamer-1} }
\begin{tabular}{ c  c c c c c c }
\hline
\hline
Directions~~&fcc & hcp\\
\hline
\hline
1~~ & 0.374 ~& 0.360 ~\\
2~~ & 0.466 ~& 0.451 ~\\
3~~ & 0.254 ~& 0.240 ~\\
\hline
\hline
\end{tabular}
\end{table}

\begin{figure}[ht]
\center{\includegraphics [width=8.0cm]{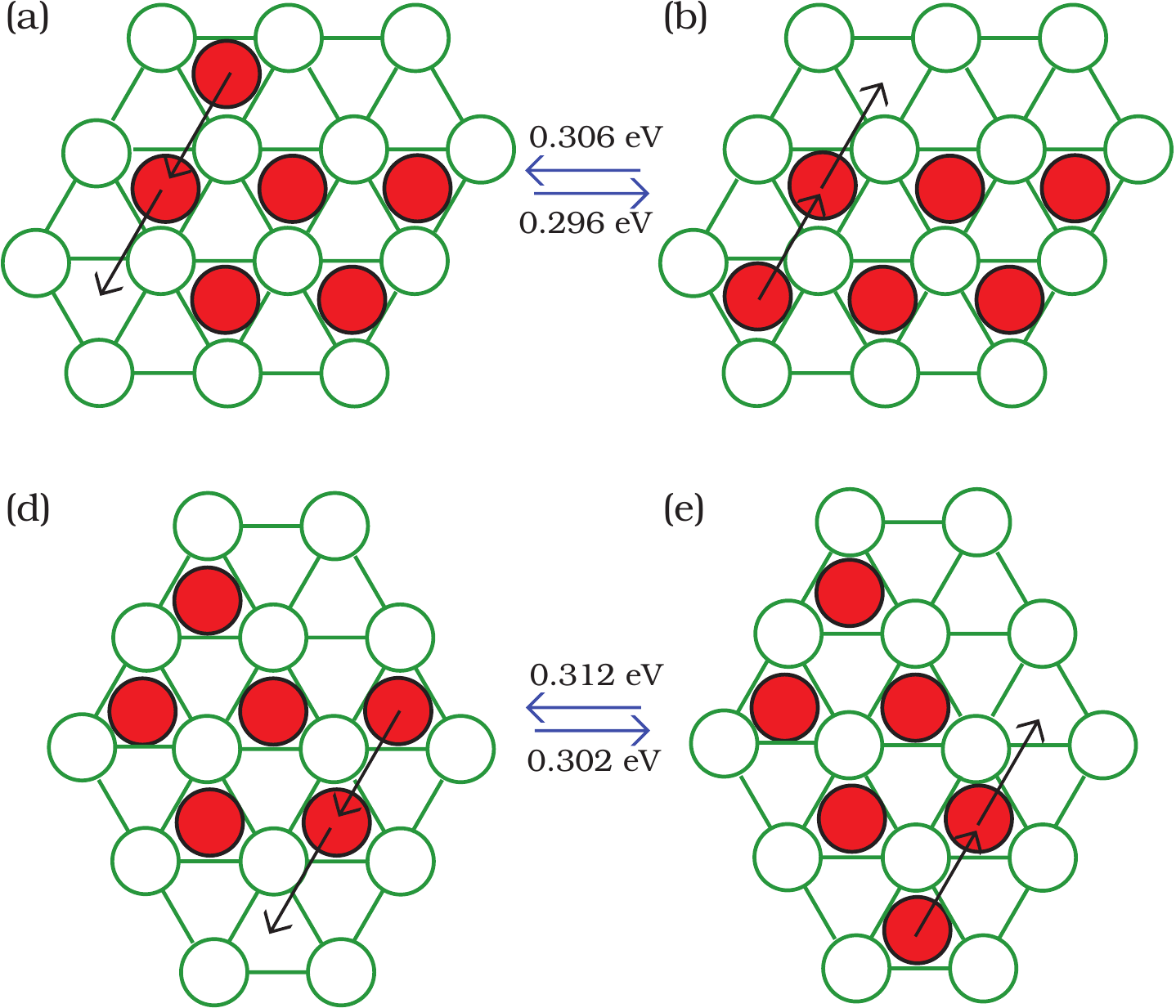} 
\caption{\label{hexamer-m}{Dimer shearing processes in case of a hexamer along with their activation barriers: (a) \& (b) all-hcp hexamers; (c) \& (d) all-fcc clusters.}}}
\end{figure}

\begin{table}[ht]
\caption{\label{thexamer-2}Activation barriers (eV) of concerted translations processes in all $3$ directions for the hexamers shown in Fig.\ref{hexamer-2} }
\begin{tabular}{ c  c c c c c c }
\hline
\hline
Directions~~&fcc & hcp\\
\hline
\hline
1~~ & 0.393 ~& 0.380 ~\\
2~~ & 0.397 ~& 0.383 ~\\
3~~ & 0.391 ~& 0.378 ~\\
\hline
\hline
\end{tabular}
\end{table}
Fig.~\ref{hexamer-m} shows the most frequently observed multi-atom processes for a hexamer -- shearing processes in which a dimer moves along the A-type step-edge of the cluster from sites of one type to the nearest-neighbor sites of the same type. Figs.~\ref{hexamer-m} (a) \& (b) show this kind of diffusion process for an hcp cluster and Figs.~\ref{hexamer-m} (c) \& (d) for an fcc cluster. Although this dimer shearing process does not much displace the center of mass of a hexamer; it does have a striking consequence: it converts a parallelogramic hexamer (Fig.~\ref{hexamer-1}) into an irregular hexamer, (Fig.~\ref{hexamer-2}) and vice-versa.

\subsection{Heptamer}

\begin{figure}
\center{\includegraphics [width=7.5cm]{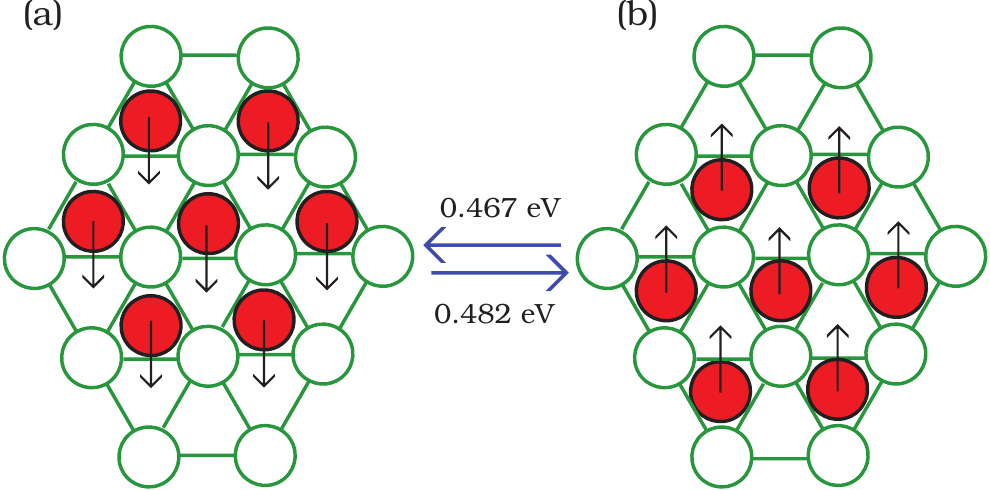} 
\caption{\label{heptamer}{Concerted diffusion processes and their activation barriers in direction $1$ for a heptamer.}}}
\end{figure}

On an fcc(111) surface, an heptamer has a compact closed-shell structure with each edge atom having at least three nearest-neighbor bonds, as shown in Fig.~\ref{heptamer}.  Our SLKMC simulations (keep in mind here the range of temperatures to which they were confined) found that heptamer diffuses exclusively via concerted diffusion processes, which displace the cluster from fcc-to-hcp and vice versa; the barriers for which are shown in Fig.~\ref{heptamer}. That these processes will predominate can also be concluded from the fact that the effective energy barrier for heptamer diffusion (cf. Table.~\ref{eff}) is close to the average of the activation barriers shown in Fig.~\ref{heptamer}. 

Since the compact heptamer has a symmetric shape, activation barriers in all three directions are the same as those shown in Fig.~\ref{heptamer}. Again, the fcc island is more energetically favorable than its hcp counterpart -- in this case by $0.015$ eV.

\subsection{Octamer}

\begin{figure}[ht]
\center{\includegraphics [width=7.0cm]{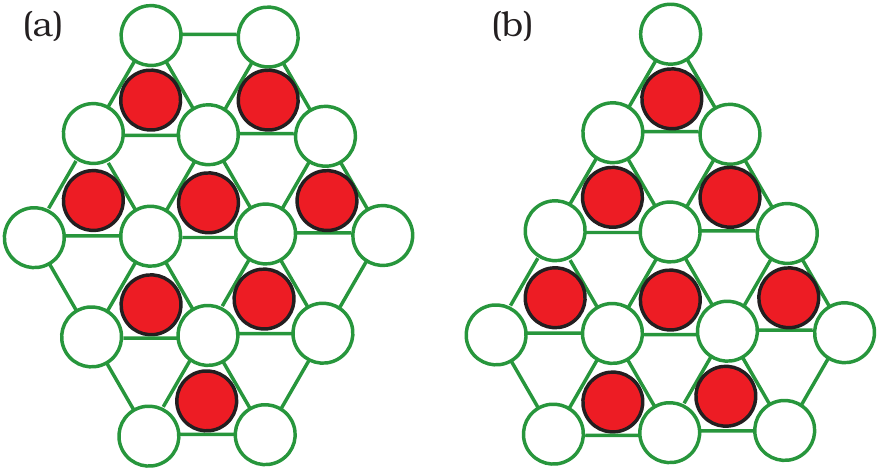} 
\caption{\label{8atom-1}{Possible orientations for a compact fcc octamer: (a) with long A-type step edges; (b) with long B-type step edges.}}}
\end{figure}
Compact octamers have two distinct orientations, one with two long A-type step-edges, the other with two long B-type step-edges, as shown in Fig.~\ref{8atom-1} (a) \& (b) for an fcc octamer. Octamers with long A-type step-edges (Fig.~\ref{8atom-1} (a)) can be obtained by attaching an atom to any B-type step-edge of a compact heptamer, while a compact octamer with long B-type step-edges results from attaching one to any A-type step-edge. Again: the fcc islands are more energetically favorable than the hcp ones, and within each type, islands with  long A-type step edges are more stable than those with long B-type step-edges.

\begin{figure}
\center{\includegraphics [width=7.5cm]{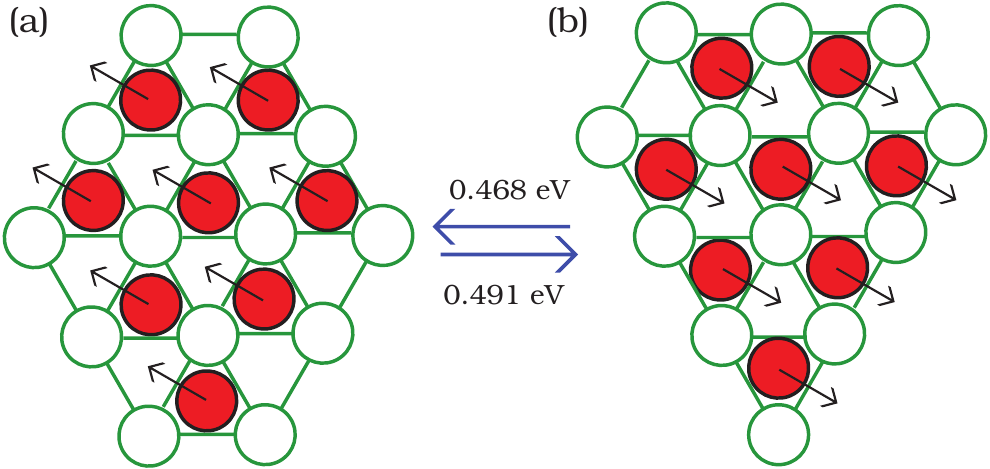} 
\caption{\label{8atm-concerted}{Examples of concerted diffusion processes in direction 2 for an octamer: (a) compact fcc octamer with long A-type step-edges; (b) compact hcp octamer with long B-type step-edges.}}}
\end{figure}

A compact octamer diffuses via concerted diffusion processes, as shown in Fig~\ref{8atm-concerted}. The activation barrier of a concerted diffusion process depends on whether the octamer has long A- or long B-type step-edges.  As Fig.~\ref{8atm-concerted} shows, a concerted diffusion process converts a long A-type step-edge fcc octamer into a long B-type step-edge hcp cluster, and vice-versa.
Table.~\ref{8atom} reports the activation barriers for concerted diffusion processes in all $3$ directions for both orientations for an fcc as well as an hcp (see Fig.~\ref{8atom-1}) octamer. Concerted diffusion processes in directions $2$ and $3$ are the most frequently observed processes in octamer diffusion.

\begin{table}
\caption{\label{8atom}Activation barriers (in eV) of the twelve concerted diffusion processes for compact octamers.}
\begin{tabular}{ c  c c c c c c }
\hline
\hline
Directions~~&fcc &hcp\\
&A~~(B)~&A (B)\\
\hline
\hline
1~~ & 0.589 (0.585) ~& 0.567 (0.571)~\\
2~~ & 0.491 (0.484)~& 0.468 (0.469)~\\
3~~ & 0.491 (0.484)~& 0.468 (0.469)~\\
\hline
\hline
\end{tabular}
\end{table}

Although an octamer diffuses primarily via concerted processes we found in our simulations that both multi-atom and single-atom processes are also relatively common. As mentioned before, we defer comprehensive discussion of single-atom processes to section ~\ref{single}. Here we discuss multi-atom processes particular to octamers.

\begin{figure}
\center{\includegraphics [width=8.5cm]{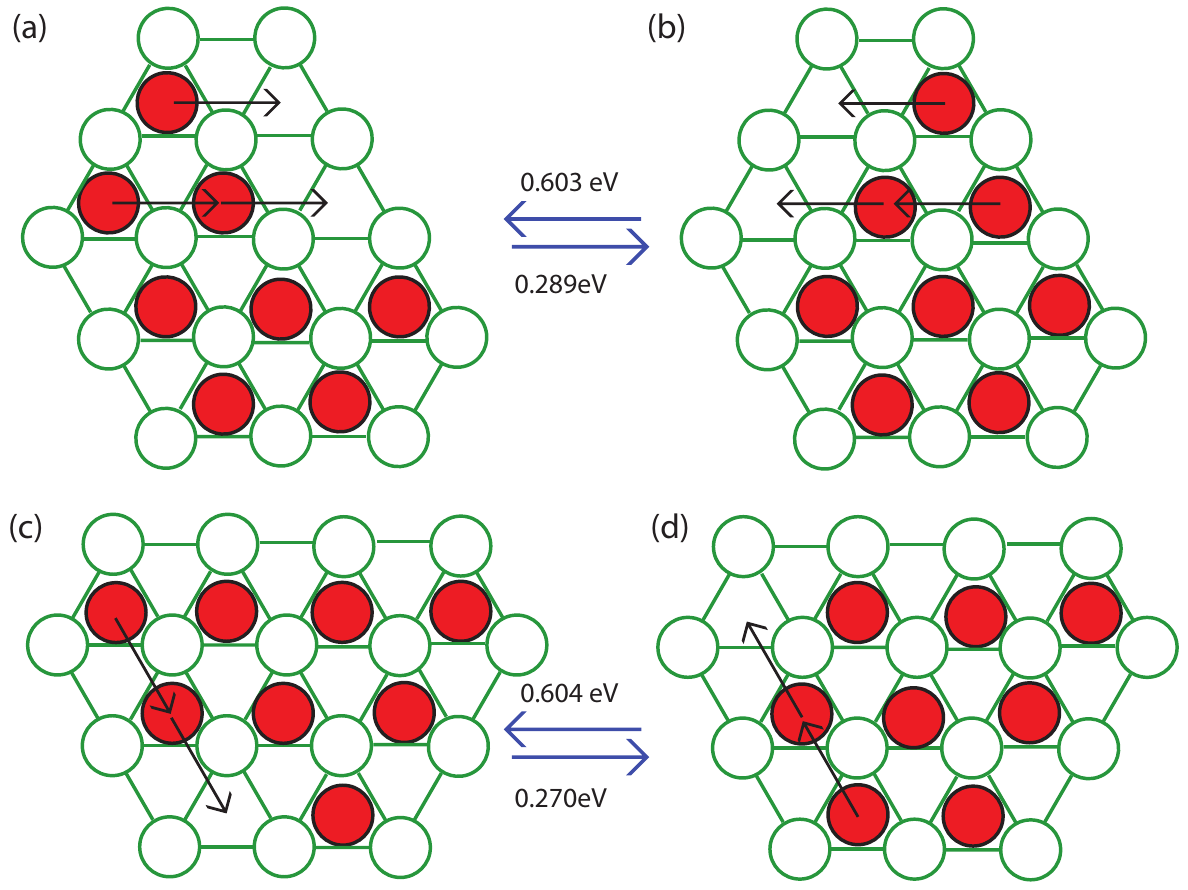} 
\caption{\label{8shearing}{Example of shearing diffusion processes within an octamer along with their activation barriers.}}}
\end{figure}
\begin{figure}                                                                                                                                                                                                                                                                                                                                                                                                                                                                                                                                                                                                                                                                                                                                                                                                                                                                                                                                                                                                                                                                                                                                                                                                                                                                                                                                                                                                                                                                                                                                                                                                                                                                                                                                                                                                                                                                                                                                                                                                                                                                                                                                                                                                                                                                                                                                                                                                                                                                                                                                                                                                                                       
\center{\includegraphics [width=8.5cm]{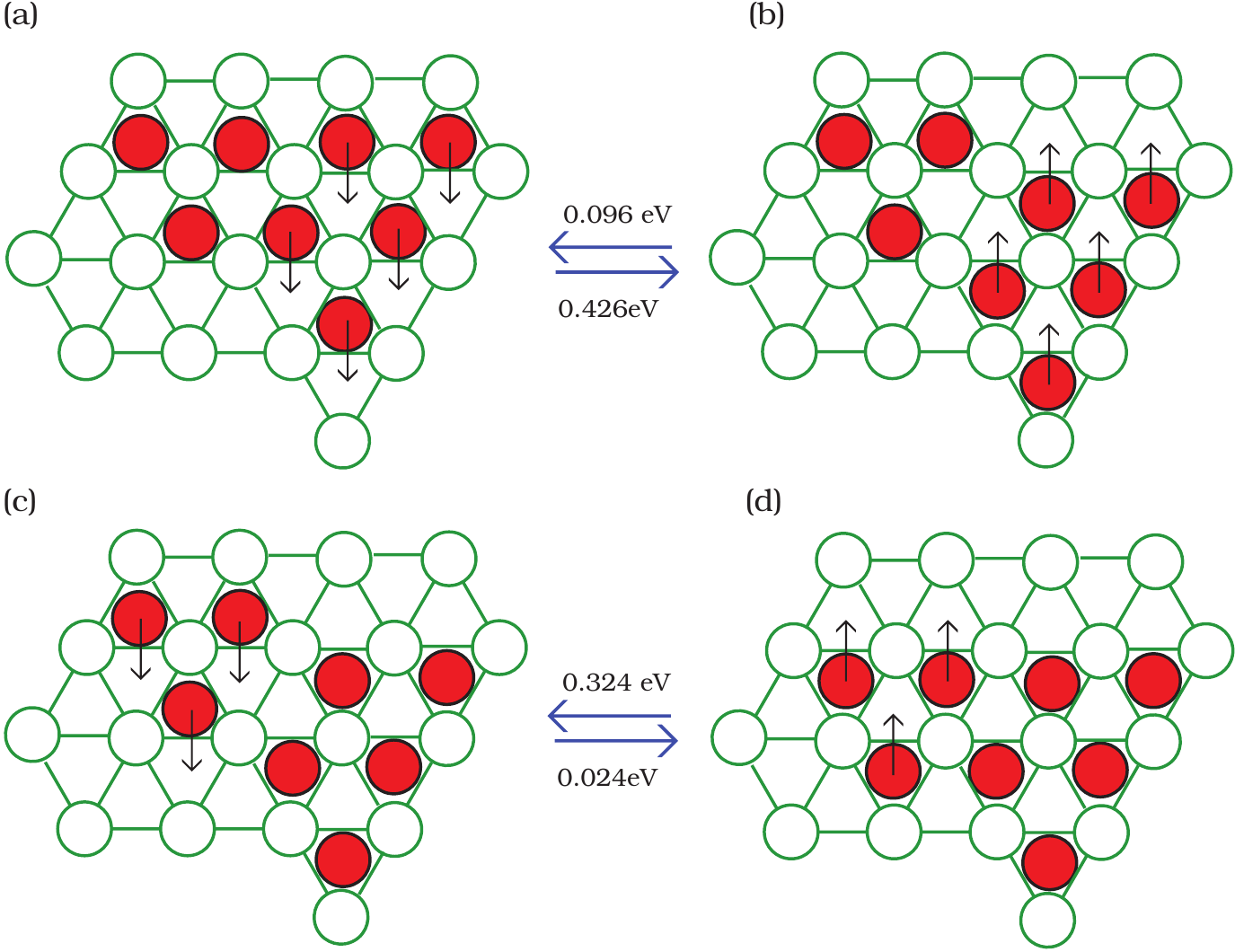} 
\caption{\label{8reptation}{Successive sub-processes (or steps) involved in an octamer reptation diffusion mechanism.}}}
\end{figure}

Multi-atom processes involving shearing and reptation are shown in Figs.~\ref{8shearing} \& \ref{8reptation}, respectively. In shearing processes, $\it part$ of the island (more than one atom) moves from fcc to the nearest fcc sites, if all the island is initially on fcc sites -- or from hcp to hcp sites, if all of it initially sits on hcp sites. Figs.~\ref{8shearing} (a) \& (b) show trimer shearing processes within an octamer, along with their activation barriers, while Figs.~\ref{8shearing} (c) \& (d) show dimer shearing processes, with their activation barriers. 

Reptation is a 2-step diffusion process. In the case of an fcc island, the entire island diffuses from fcc to nearest-neighbor hcp sites in two steps. In the first, part of the island moves from fcc to nearest-neighbor hcp sites, leaving part of the island on fcc sites and part on hcp sites. In the next, the remainder of the island initially on fcc sites moves to hcp sites. Figs.~\ref{8reptation} (a)-(d) shows various steps (subprocesses) of a reptation process, with their activation barriers.

\subsection{Nonamer}

\begin{figure}[ht]
\center{\includegraphics [width=8.5cm]{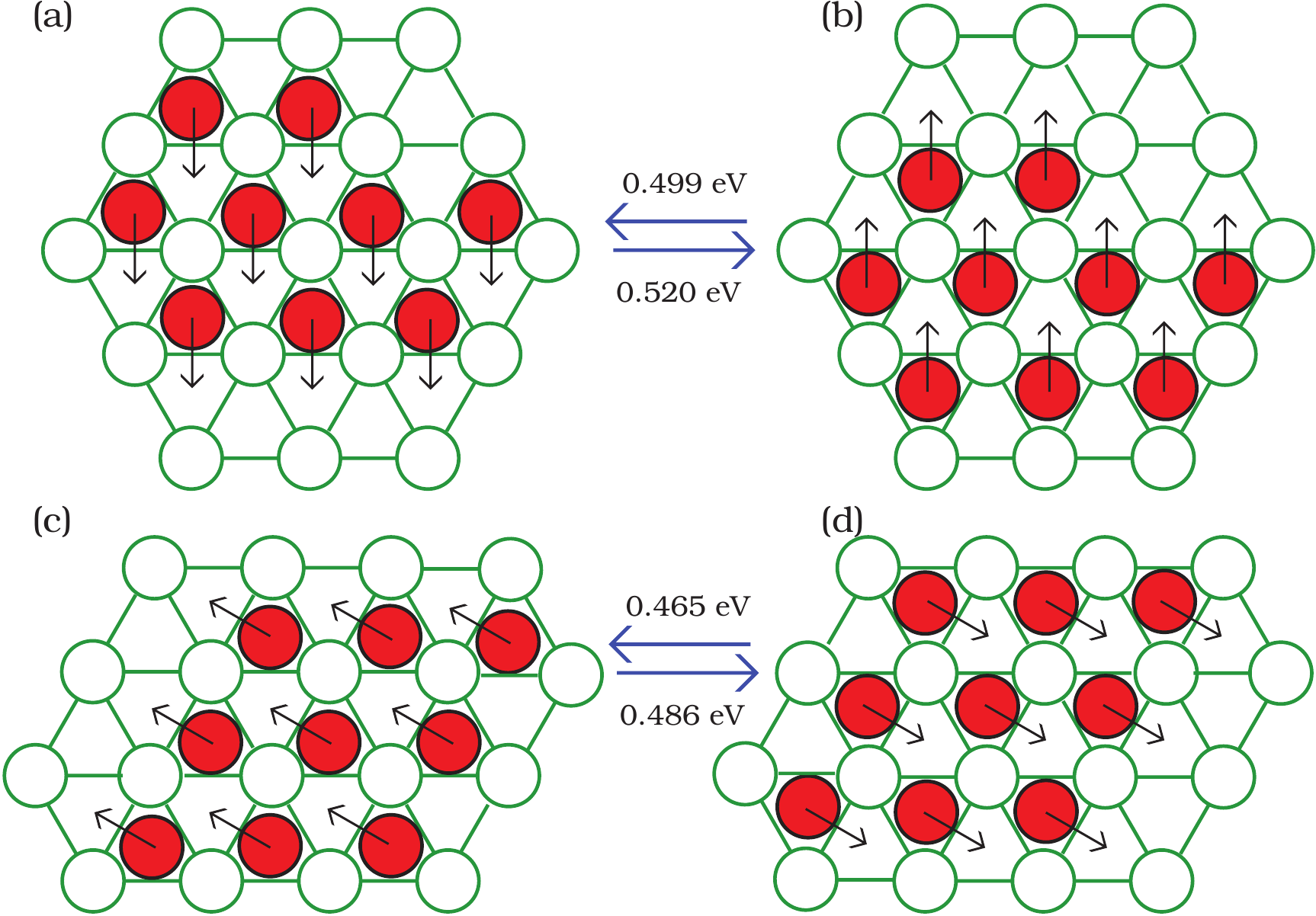} 
\caption{\label{9concerted}{Concerted diffusion processes and their activation barriers for nonamers.}}}
\end{figure}

\begin{figure}[ht]
\center{\includegraphics [width=8.0cm]{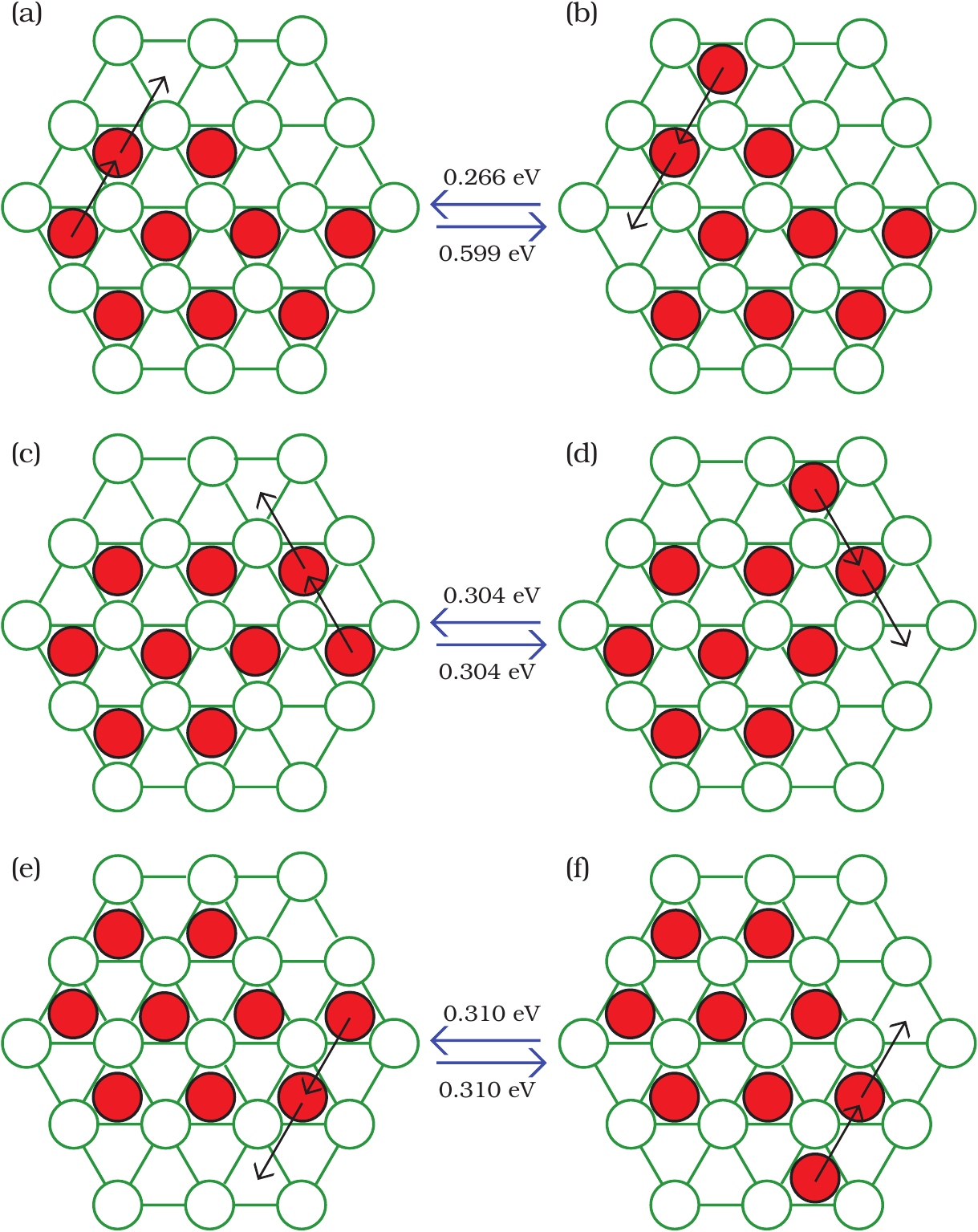} 
\caption{\label{9multi}{Dimer shearing processes and their activation barriers for compact nonamers.}}}
\end{figure}
For a nonamer, we observed all types of diffusion processes -- single-atom, multi-atom and concerted. The most frequently observed were two types of single-atom mechanisms: edge-diffusion processes along an A- or a B-type step-edge and corner rounding (Fig.~\ref{psingle}). The nonamer is the smallest island for which concerted processes are not the most frequently picked (the concerted processes for the most frequently observed nonamer configurations [compact or nearly so] are shown in Fig.~\ref{9concerted}). Even so, concerted processes contribute the most to $\it island$ diffusion: that is, the displacement they produce in the nonamer's center of mass is far greater than that produced by single-atom processes, despite the far greater frequency of the latter. This is reflected in the fact that the effective energy barrier for nonamer (cf. Table~\ref{teff}) is much closer to the average activation barrier for concerted processes (cf. Table.~\ref{t9concerted}) than for that of single-atom processes (cf. Table~\ref{t9concerted}). The fact that the effective activation barrier is slightly higher than the average energy barrier for concerted processes is due mainly to the contribution of kink processes, which do contribute somewhat to island diffusion.
\begin{table}[ht]
\caption{\label{t9concerted}Activation barriers (eV) of concerted translations processes in all $3$ directions for the nonamers shown in Figs.\ref{9concerted}(a--d).}
\begin{tabular}{ c  c c c }
\hline
\hline
Directions~~&fcc&  hcp\\
~~&(a)&  (b)\\
\hline
\hline
1~~ & 0.520 ~& 0.499 ~\\
2~~ & 0.626 ~& 0.605 ~\\
3~~ & 0.605 ~& 0.583 ~\\
\hline
\hline
\end{tabular}
\begin{tabular}{ c  c c c  }
\hline
\hline
fcc &  hcp\\
(c)& (d)\\
\hline
\hline
0.486 ~& 0.465 ~\\
0.486 ~& 0.465 ~\\
0.693 ~&0.672~\\
\hline
\hline
\end{tabular}
\end{table}

The most frequently observed multi-atom processes are the four forms of dimer shearing along an A-type step-edge shown in Fig.~\ref{9multi}(c)-(f), as have been discussed above for island of sizes 6 \& 8. The activation barriers for these dimer shearing processes are lower than those for single-atom diffusion processes along an edge and also for some corner rounding processes. Reptation processes also show up, but only when the nonamer is non-compact (we do not illustrate these here)~\cite{slkmcII}.
\subsection{Decamer}
\begin{figure}[ht]
\center{\includegraphics [width=8.5cm]{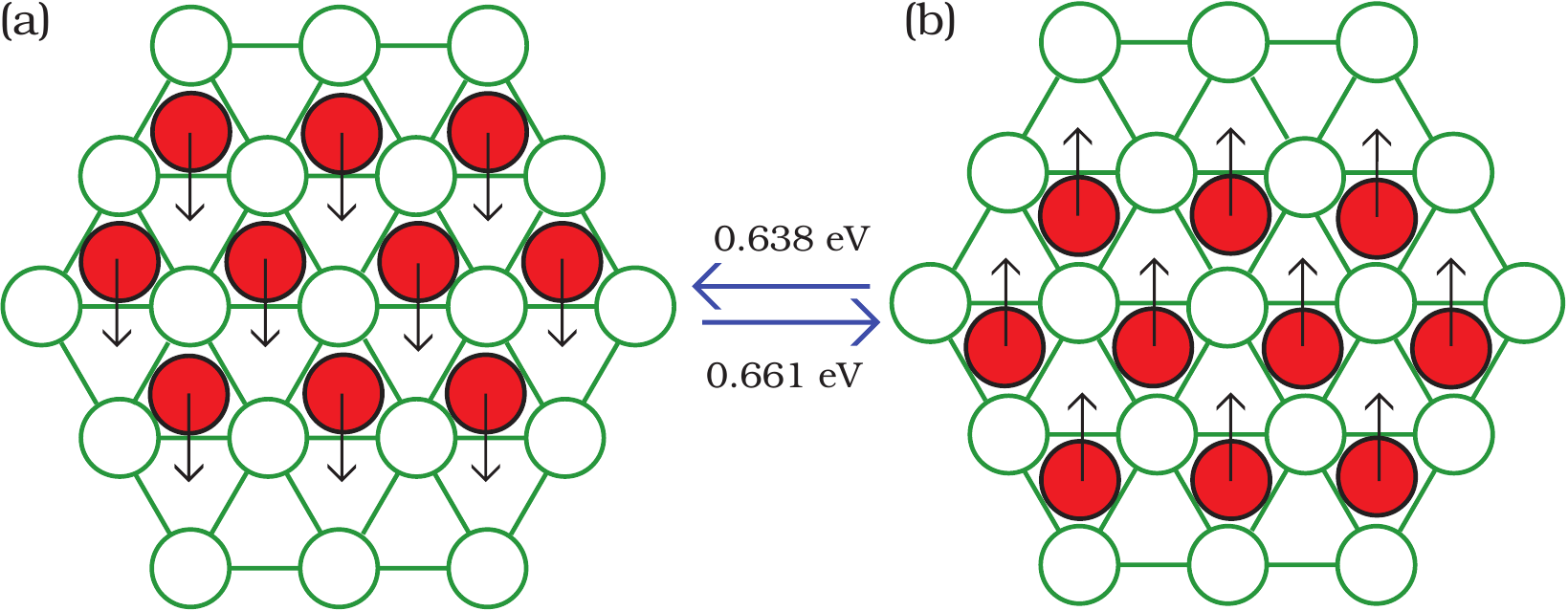} 
\caption{\label{10concerted}{Frequent concerted diffusion processes and their activation barriers for compact decamers.}}}
\end{figure}

Even in the case of a decamer, we have observed single-atom, multi-atom and concerted diffusion processes. Single-atom diffusion processes are the most frequently observed. The most frequently observed compact shape of decamer during our simulations is that shape shown in Fig.~\ref{10concerted}, which has the same number of A- and B-type step edges. As usual, an fcc cluster is energetically more favorable than an hcp cluster. 
\begin{table}[ht]
\caption{\label{t10concerted}Activation barriers (eV) of concerted diffusion processes in all $3$ directions of decamer as shown in Fig.\ref{10concerted} (a) \& (b).}
\begin{tabular}{ c  c c c c c c }
\hline
\hline
Directions~~&fcc &  hcp\\
\hline
\hline
1~~ & 0.661 ~& 0.638 ~\\
2~~ & 0.700 ~& 0.677~\\
3~~ & 0.700 ~& 0.677 ~\\
\hline
\hline
\end{tabular}
\end{table}

For the shape shown in Fig.~\ref{10concerted}, the most frequently observed concerted diffusion processes are those shown in the same figure, along with their activation barriers reported in Table~\ref{t10concerted}. It can be seen from the Table~\ref{eff}, that the effective energy for decamer diffusion is close to that of the average energy barrier of these concerted processes. That is why decamer diffusion is dominated by concerted processes.
\begin{figure}[ht]
\center{\includegraphics [width=8.0cm]{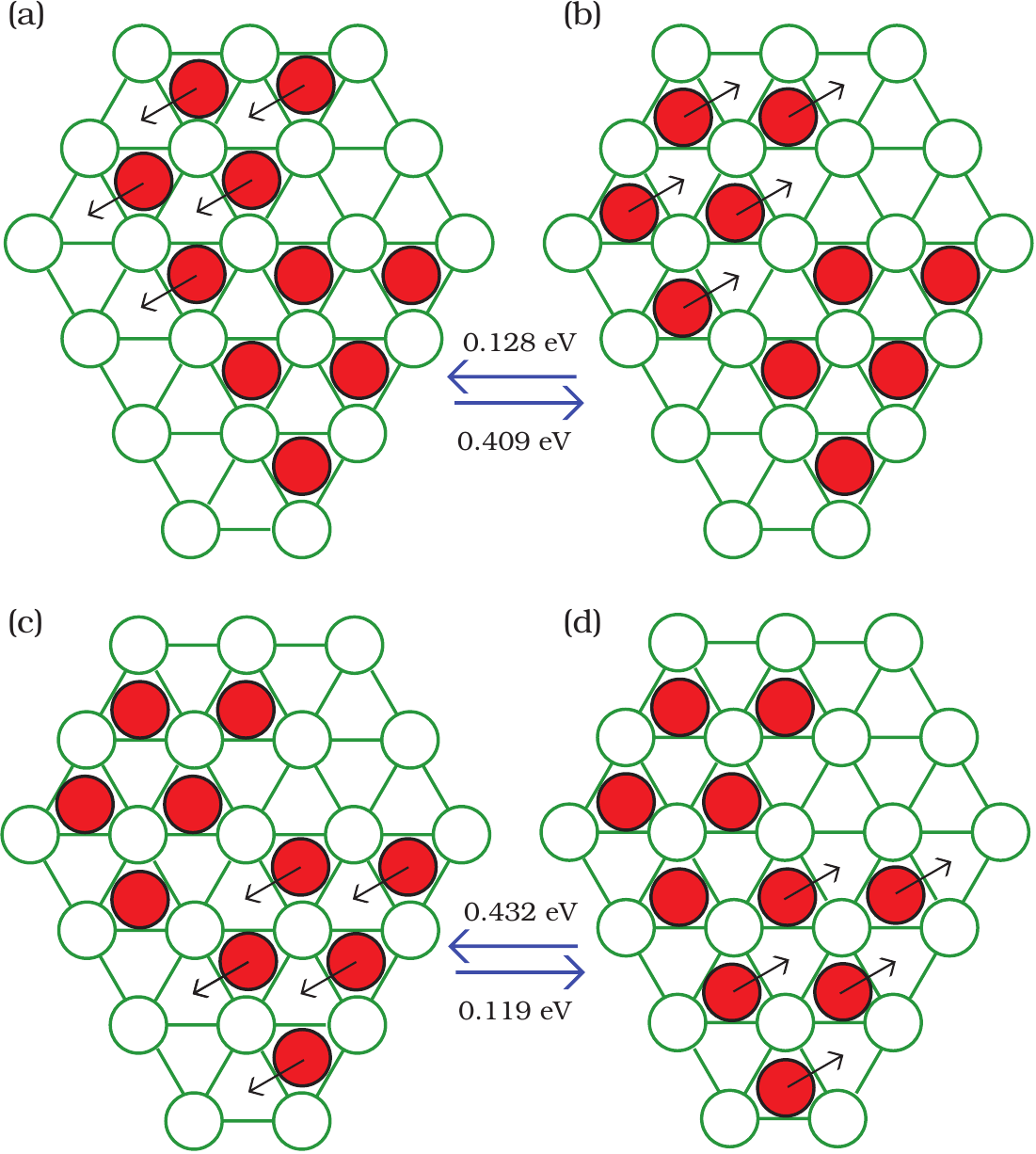} 
\caption{\label{10reptation}{Diffusion steps (sub-processes) in decamer reptation. Note that the decamers in (b) and (c) are identical, though the arrows indicate different processes.}}}
\end{figure}

As with the nonamer, a decamer also undergoes multi-atom processes (shearing and reptation). Of these, the most frequently observed is dimer shearing along an A-type step-edge, similar to what has been discussed for clusters of size 6, 8 and 9. Fig.~\ref{10reptation} shows the sub-processes in a reptation process, along with the activation barrier of each.

\subsection{Single-atom Processes}\label{single}

\begin{figure}
\center{\includegraphics [width=8.5cm]{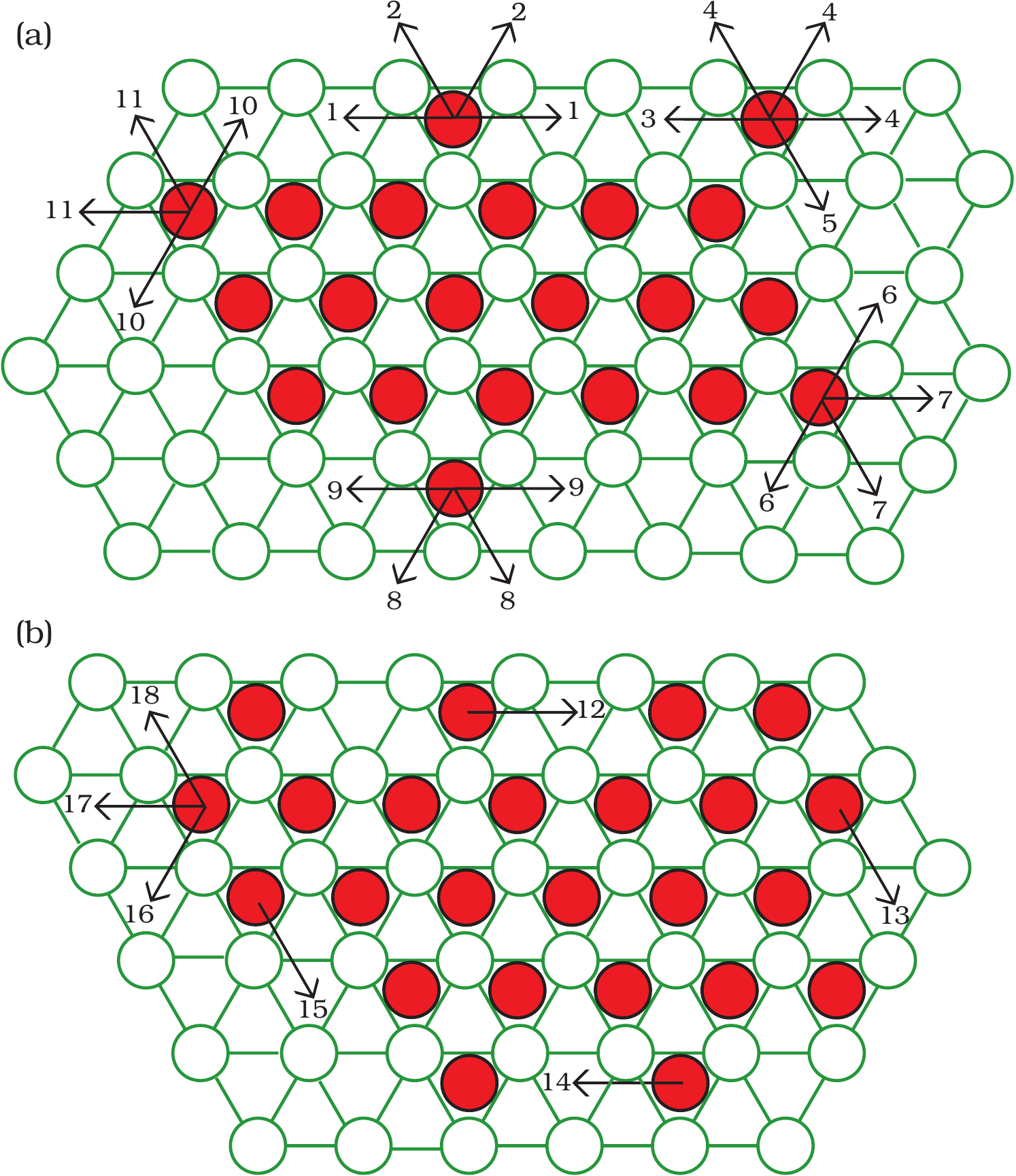} 
\caption{\label{psingle}{Single-atom processes for an hcp island (Though analogous processes occur for an fcc island, we do not illustrate them here). The index numbers designate the processes described in Table~\ref{tsingle}, which gives the activation barriers for each.}}}
\end{figure}

In this section we provide detail about single-atom processes: edge-diffusion, corner rounding, kink attachment, and kink detachment, as shown in Fig.~\ref{psingle} for an hcp island. Their corresponding activation barriers and those for their fcc analogues are given in Table~\ref{tsingle}. In each single-atom process, an atom on an fcc site moves to a nearest-neighbor fcc site, while an atom on hcp site moves to a nearest-neighbor hcp site. The activation barriers for single-atom processes depend not only on whether the atom is part of an fcc sland or an hcp island, but also on whether the diffusing atom is on an A-type or a B-type step-edge.

In classifying single-atom processes in Table~\ref{tsingle} we have used the notation X$_{n_{i}}$U $\rightarrow$ Y$_{n_{f}}$V, where  where X or Y = A (for an A-type step-edge) or B (for a B-type step-edge) or K (for kink) or C (for corner) or M (for monomer); n$_{i}$ = the number of nearest-neighbors of the diffusing atom before the process; n$_{f}$ = the number of that atom's nearest neighbors after the process. U or V = A or B (for corner or kink processes) or null (for all other other process types).

For example, process 1, B$_{2}$ $\rightarrow$ B$_{2}$, is a single-atom B-step edge process in which the diffusing atom has 2 nearest-neighbors before and after the process. Process 3, C$_{1}$B $\rightarrow$ B$_{2}$, is a corner rounding process towards a B-step, the diffusing atom starting on the corner of a B-step with one nearest-neighbor and ending up on the B-step with two nearest-neighbors. In process 10, C$_{2}$A $\rightarrow$ C$_{1}$B, the diffusing atom begins on the corner of an A-step having two nearest-neighbors and ends up on the corner of a B-step with only one nearest-neighbor.

\begin{table}
\caption{\label{tsingle}Activation barriers (eV) of single-atom processes for both fcc and hcp islands. The index numbers refer to the types of processes illustrated in Fig.~\ref{psingle}. See text for explaination of the notation used to classify the process types.}
\begin{tabular}{ c  l c c}
\hline
\hline
Index no. ~&Process type&fcc & hcp\\
\hline
\hline
1~~ &B$_{2}$ $\rightarrow$ B$_{2}$& 0.454 ~& 0.448 ~\\
2~~ & B$_{2}$ $\rightarrow$ M&0.821 ~& 0.815~\\
3~~ &C$_{1}$B $\rightarrow$ B$_{2}$ &0.177 ~& 0.173 ~\\
4~~ &C$_{1}$B $\rightarrow$ M &0.458 ~& 0.455 ~\\
5~~ &C$_{1}$B $\rightarrow$ A$_{2}$ &0.040 ~& 0.038 ~\\
6~~ &C$_{1}$B $\rightarrow$ C$_{1}$B &0.540 ~& 0.585 ~\\
7~~ &C$_{1}$B $\rightarrow$ M &0.811 ~& 0.809 ~\\
8~~ &A$_{2}$ $\rightarrow$ M &0.795 ~& 0.794 ~\\
9~~ &A$_{2}$ $\rightarrow$ A$_{2}$ &0.326 ~& 0.307 ~\\
10~~ &C$_{2}$A $\rightarrow$ C$_{1}$B &0.399 ~&  0.397~\\
11~~ &C$_{2}$A $\rightarrow$ M &0.787 ~& 0.785 ~\\
12~~ &B$_{2}$ $\rightarrow$ K$_{3}$B &0.415 ~& 0.298 ~\\
13~~ &K$_{3}$A $\rightarrow$ A$_{2}$ &0.601 ~& 0.701 ~\\
14~~ &A$_{2}$ $\rightarrow$ K$_{3}$A &0.302 ~& 0.389 ~\\
15~~ &K$_{3}$B $\rightarrow$ B$_{2}$ &0.729 ~& 0.597 ~\\
16~~ &K$_{3}$B $\rightarrow$ C$_{1}$B &0.731 ~& 0.787 ~\\
17~~ &K$_{3}$B $\rightarrow$ M &1.138 ~& 1.150 ~\\
18~~ &K$_{3}$B $\rightarrow$ C$_{1}$B &0.820 ~& 0.759 ~\\
\hline
\hline
\end{tabular}
\end{table}
\section{Diffusion coefficients and effective energy barriers}\label{Dcoefficients}
We start our SLKMC simulations with an empty database. Every time a new configuration (or neighborhood) is turned up, SLKMC-II finds on the fly all possible processes using the drag method, calculates their activation barriers and stores them in the database as the simulation proceeds. Calculation of energetics occurs at each KMC step during initial stages of the simulation when the database is empty or nearly so, and ever less frequently later on. Recall that the types of processes and their activation barriers are dependent on island size, each one of which requires a separate database that cannot be derived from that for islands of other sizes.
\begin{figure}[ht]
\center{\includegraphics [width=7.0cm]{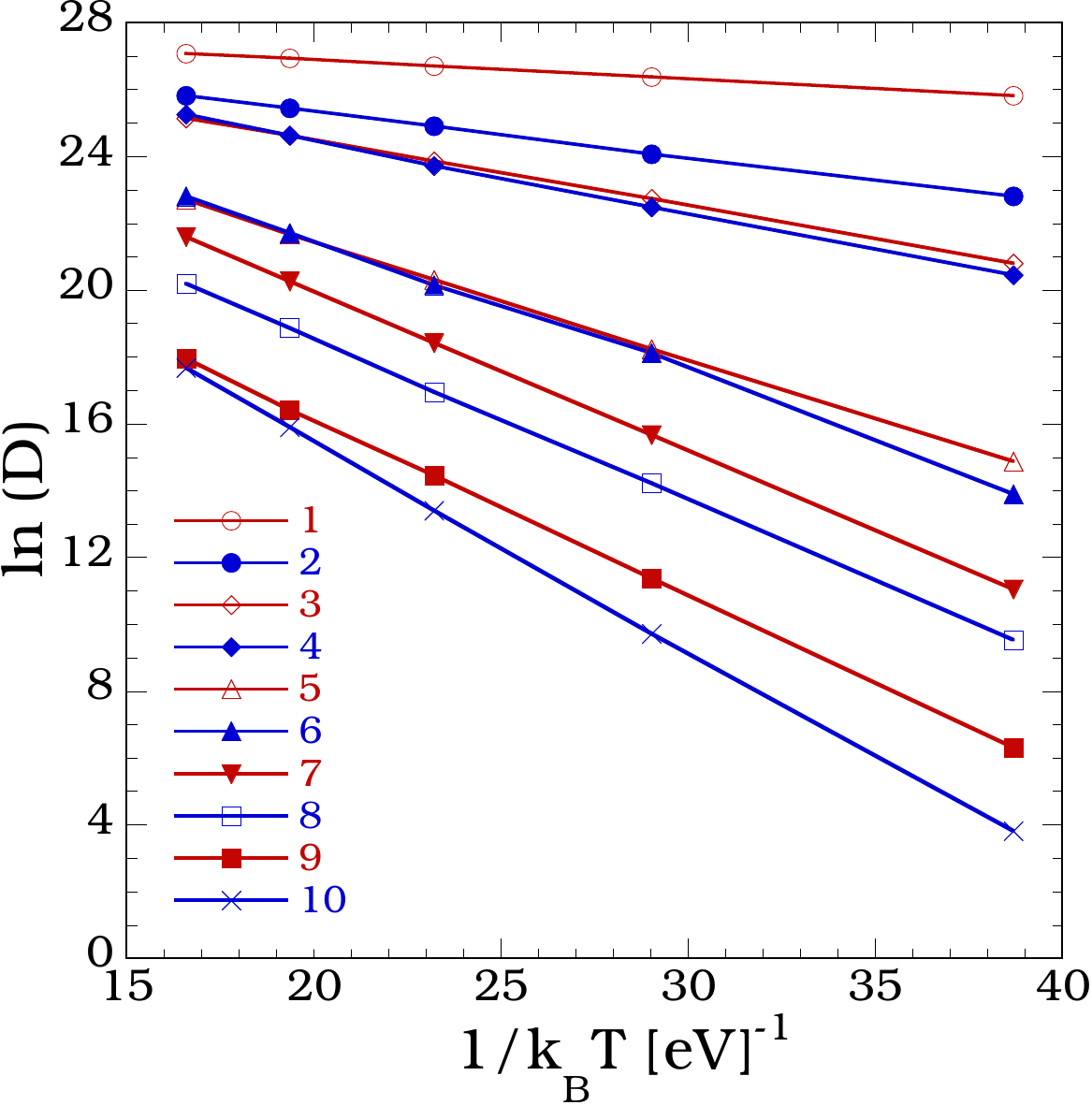} 
\caption{\label{arr}{Arrhenious plots for 1$\--$10 atom islands.}}}
\end{figure}

\begin{figure}[ht]
\center{\includegraphics [width=7.0cm]{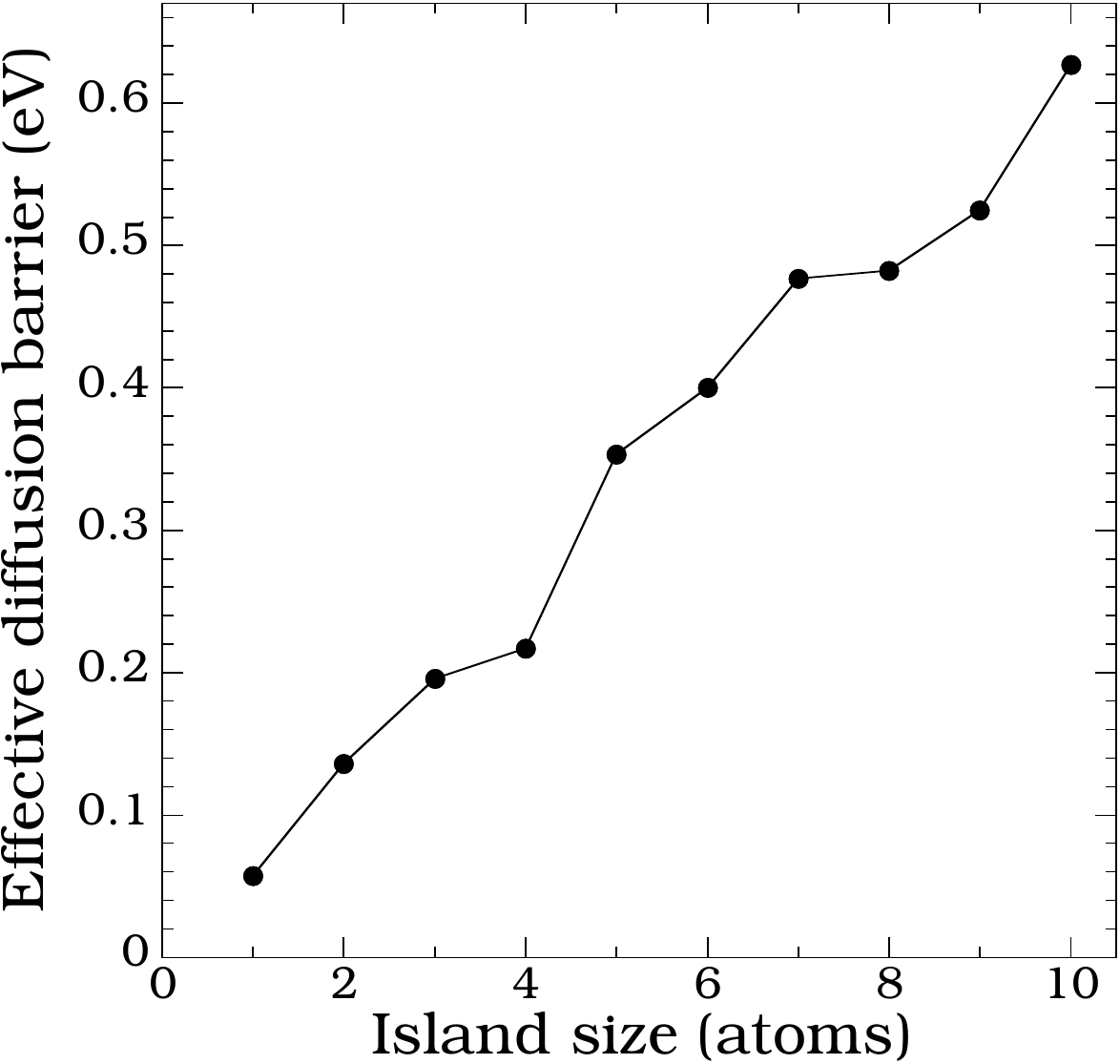} 
\caption{\label{eff}{Effective energy barriers of 1$\--$10 atom islands as a function of island size.}}}
\end{figure}

We carried out 10$^{7}$ KMC steps for each island size at temperatures 300K, 400K, 500K, 600K and 700K. We calculated the diffusion coefficient of an island of a given size using Einstein Equation:\cite{MSD} $D = \lim_{t \to \infty} \langle R_{CM}(t) - R_{CM}(0)]^{2}\rangle/2dt$, where $R_{CM}(t)$ is the position of the center of mass of the island at time $t$, and  $d$ is the dimensionality of the system, which in our case is $2$. Diffusion coefficients thus obtained for island sizes $1-10$ at various temperatures are summarized in Table~\ref{teff}. At $300$ K, diffusion coefficients range from $1.63\times10^{11}$ \AA$^{2}$/s for a monomer to $8.66 \times10^{01}$ \AA$^{2}$/s for a decamer.
Effective energy barriers for islands are extracted from their respective Arrhenius plots (Fig.~\ref{arr}) and also summarized in Table~\ref{teff}. Fig.~\ref{eff} plots effective energy barrier as a function of island size. It can be seen that the effective energy barrier increases almost linearly with island size. 

\begin{table*}[ht]
\caption{\label{teff}Diffusion coefficients ((\AA$^{2}$/s) at various temperatures and effective energy barriers for Ni islands.}
\begin{tabular}{ c | c c c c c c }
\hline
\hline
Island Size &300K & 400K & 500K & 600K& 700K & $E_{eff}$ (eV)\\
\hline
\hline
1 & $1.63\times 10^{11}$~ & $2.85\times 10^{11}$ ~& $3.99\times 10^{11}$~& $5.00\times 10^{11}$~ & $5.87\times 10^{11}$ ~& 0.058 \\ 
2 & $8.14\times 10^{09}$ ~& $2.88\times 10^{10}$~& $6.54\times 10^{10}$~ & $1.13\times 10^{11}$ ~& $1.64\times 10^{11}$ ~& 0.136\\
3 & $1.09\times 10^{09}$ ~& $7.42\times 10^{09}$~& $2.30\times 10^{10}$~ & $5.00\times 10^{10}$ ~& $8.23\times 10^{10}$ ~& 0.196\\
4 & $7.64\times 10^{08}$ ~& $5.83\times 10^{09}$~& $2.02\times 10^{10}$~ & $4.91\times 10^{10}$ ~& $9.93\times 10^{10}$ ~& 0.217\\
5 & $2.90\times 10^{06}$ ~& $8.42\times 10^{07}$~& $6.59\times 10^{08}$~ & $2.55\times 10^{09}$ ~& $7.36\times 10^{09}$ ~& 0.353\\
6 & $1.08\times 10^{06}$ ~& $7.39\times 10^{07}$~& $5.66\times 10^{08}$~ & $2.69\times 10^{09}$ ~& $8.12\times 10^{09}$ ~& 0.400\\
7 & $6.24\times 10^{04}$ ~& $6.40\times 10^{06}$~& $1.01\times 10^{08}$~ & $6.37\times 10^{08}$ ~& $2.37\times 10^{09}$ ~& 0.477\\
8 & $1.38\times 10^{04}$ ~& $1.50\times 10^{06}$~& $2.31\times 10^{07}$~ & $1.57\times 10^{08}$ ~& $5.91\times 10^{08}$ ~& 0.482\\
9 & $5.48\times 10^{02}$ ~& $8.77\times 10^{04}$~& $1.89\times 10^{06}$~ & $1.35\times 10^{07}$ ~& $6.23\times 10^{07}$ ~& 0.525\\
10 & $8.66\times 10^{01}$ ~& $2.25\times 10^{04}$~& $7.20\times 10^{05}$~ & $7.97\times 10^{06}$ ~& $4.95\times 10^{07}$ ~& 0.627\\
\hline
\hline
\end{tabular}
\end{table*}
\section{Further Discussion And Conclusions}\label{discussion}
To summarize: we have performed a systematic study of the diffusion of small Ni islands (1-10 atoms) on Ni(111), using a self-learning KMC method with a newly-developed pattern recognition scheme (SLKMC-II) in which the system is allowed to evolve through mechanisms of its choice on the basis of a self-generated database of single-atom, multiple-atom and concerted diffusion processes (each with its particular activation barrier) involving fcc-fcc, fcc-hcp and hcp-hcp jumps. We find that concerted diffusion processes contribute the most to the displacement of the center of mass (i.e., to island diffusion), while single-atom processes contribute the least. As for multi-atom processes (reptation or shearing): while these produce more displacement than the latter, they are hardly ever selected. Though the energy barriers for reptation processes are small compared to those for concerted diffusion processes, reptation occurs only when an island is transformed into a non-compact shape, as happens only rarely in the temprature range to which our study is confined. In contrast, though shearing occurs with close-to-compact shapes (which appear more frequently than do non-compact shapes, but with islands of certain sizes), the barriers for these processes are higher than those for reptation. Finally, although for all island sizes, island diffusion is primarily dominated by concerted diffusion processes, the frequency of occurrence of both single-atom and multi-atom processes does increase with increase in island size, owing to increase in the activation barrier for concerted diffusion processes with island size. 

\begin{figure}
\center{\includegraphics [width=7.0cm]{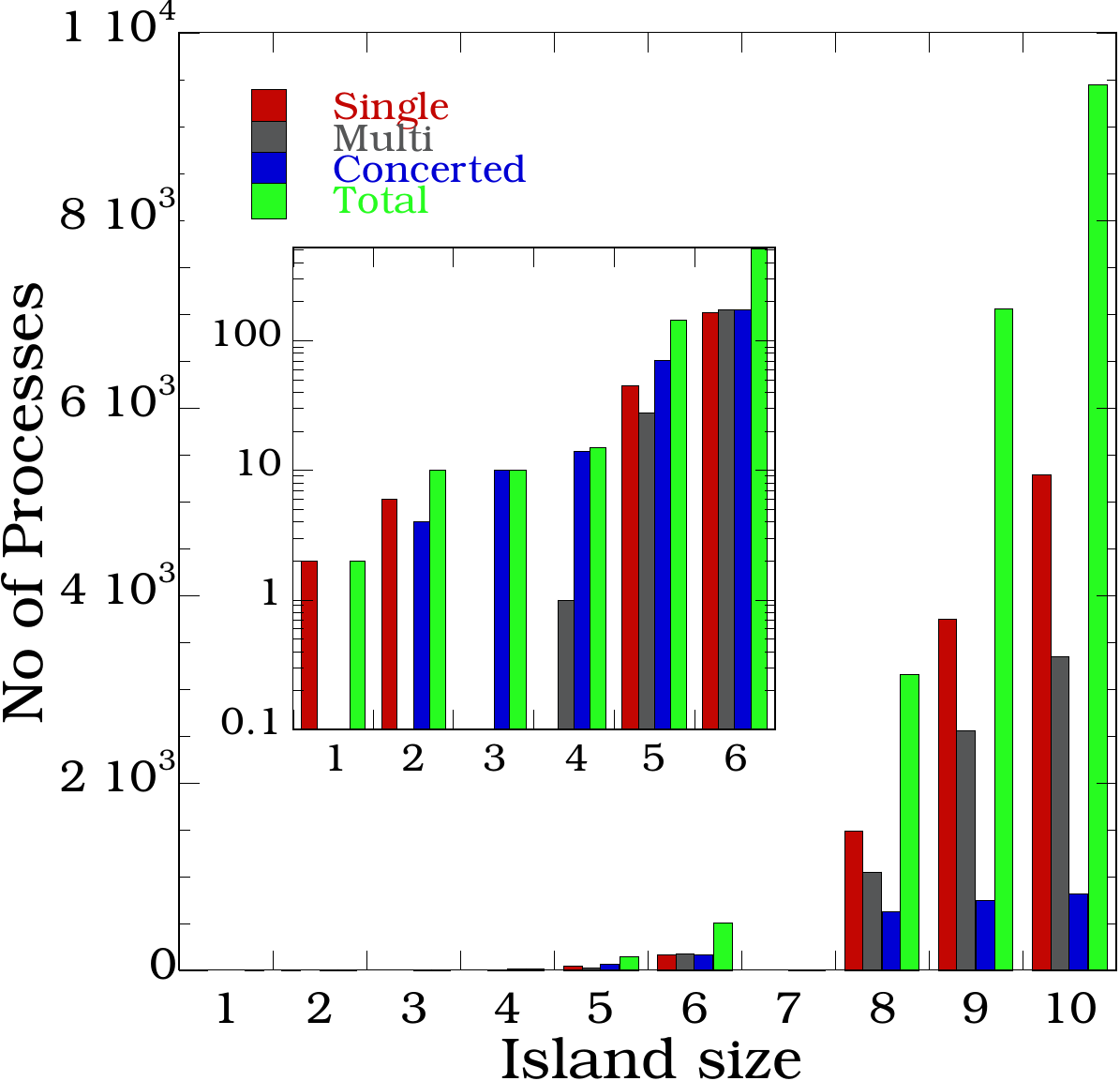} 
\caption{\label{database}{Distribution of single-atom, multi-atom, concerted and total processes for 1-10 atom islands accumulated in the database during SLKMC simulations. Inset shows the log-linear plot for up to the 6-atom island.}}}
\end{figure}

Fig.~\ref{database} shows the number of each type of diffusion process (single-, multi-atom and concerted) collected during our SLKMC-II simulations for each island size (1-10), together with the total number of processes of all types for each island size. (For the sake of clarity, the insert shows a log-linear plot of these quantities for island sizes 1-6). It can be seen that the number of processes accumulated increases with island size, and significantly so beyond the tetramer. It can also be seen from Fig.~\ref{database} that the overall increase in number of processes with island size is constituted predominantly by significant increases in single-atom processes and (to a lesser degree) multi-atom processes.
Meanwhile the number of concerted processes accumulated in the database increases at a much slower pace with island size. 

This significant increase in single-atom processes is mainly due to the use of $10$ rings to identify the neighborhood of an atom. Use of $10$ rings corresponds to inclusion of $5$ nearest-neighbor fcc-fcc or hcp-hcp interactions. Elsewhere we show that $6$ rings (which corresponds to $3$ nearest-neighbor interactions) offer a range of interaction sufficient for accurately calculating the activation barriers for single-atom processes.\cite{tobepublished}   But it is essential to include the long-range interaction (and hence 10 rings) if one aims to accurately take into account multi-atom and concerted processes, the latter of which predominate in small-island diffusion.  This significant increase in the number of processes with island size also justifies resorting to an automatic way of finding all the possible processes during simulations instead of using a fixed (and thus necessarily preconceived) list of events.

As mentioned earlier,  with increasing island size not only does the number of accumulated single-atom processes and multi-atom processes increase but also their frequency of occurrence. Still island diffusion is primarily due to concerted diffusion processes, since it is these that produce largest displacement of center of mass. This can be easily observed by comparing, for each island size, the effective diffusion barriers given in Table~\ref{teff} with the activation barriers in the tables given in Sect.~\ref{results} for concerted diffusion processes: effective diffusion barriers more or less closely follow activation barriers for concerted diffusion processes -- except for the 9-atom island, in which the contribution of single-atom processes to the island's diffusion is significantly larger than for other island sizes. 

In conclusion, Ni small-island diffusion on Ni (111) is primarily due to to concerted diffusion processes, even though the frequency of their occurrence decreases with increase in island size.

\begin{acknowledgments}

We would like to acknowledge computational resources provided by University of Central Florida. We also thank Lyman Baker for critical reading of the manuscript and Oleg Trushin for his suggestions during initial stages of this work.

\end{acknowledgments}

 \end{document}